\documentclass[conference]{IEEEtran}
\usepackage{blindtext, graphicx}
\usepackage{listings}
\usepackage{tikz}
\usepackage{amsmath,amssymb,amsfonts}
\usepackage{subfigure}
\usepackage{textcomp}
\usepackage{listings}
\usepackage{lscape}
\usepackage{array}
\usepackage{url}
\usepackage{longtable}
\usepackage{algpseudocode}
\usepackage{enumerate}
\usepackage{booktabs}
\usepackage{multirow}
\usepackage{multicol}
\usepackage{lscape}
\usepackage{algorithm}
\usepackage[utf8]{inputenc}
\usepackage[
    backend=biber, 
    natbib=true,
    style=numeric,
    sorting=none
]{biblatex}
\definecolor{codegreen}{rgb}{0,0.6,0}
\definecolor{codegray}{rgb}{0.5,0.5,0.5}
\definecolor{codepurple}{rgb}{0.58,0,0.82}
\definecolor{backcolour}{rgb}{0.95,0.95,0.92}

\lstdefinestyle{mystyle}{
    backgroundcolor=\color{backcolour},   
    commentstyle=\color{codegreen},
    keywordstyle=\color{magenta},
    numberstyle=\tiny\color{codegray},
    stringstyle=\color{codepurple},
    basicstyle=\ttfamily\footnotesize,
    breakatwhitespace=false,         
    breaklines=true,                 
    captionpos=b,                    
    keepspaces=true,                 
    numbers=left,                    
    numbersep=5pt,                  
    showspaces=false,                
    showstringspaces=false,
    showtabs=false,                  
    tabsize=2
}

\def\BibTeX{{\rm B\kern-.05em{\sc i\kern-.025em b}\kern-.08em
    T\kern-.1667em\lower.7ex\hbox{E}\kern-.125emX}}

\title{Navigating the Concurrency Landscape: A Survey of Race Condition Vulnerability Detectors}

\author{
\IEEEauthorblockN{Aishwarya Upadhyay, Vijay Laxmi, Smita Naval}
\IEEEauthorblockA{Department of Computer Science and Engineering\\
Malaviya Naional Institue of Technology Jaipur \\ Email: 2022rcp9026@mnit.ac.in, vlaxmi@mnit.ac.in, smita.cse@mnit.ac.in}
}

\begin{document}

\maketitle
\thispagestyle{empty}
\pagestyle{empty}

\begin{abstract}
As technology continues to advance and we usher in the era of Industry 5.0, there has been a profound paradigm shift in operating systems, file systems, web, and network applications. The conventional utilization of multiprocessing and multicore systems has made concurrent programming increasingly pervasive. However, this transformation has brought about a new set of issues known as concurrency bugs, which, due to their wide prevalence in concurrent programs, have led to severe failures and potential security exploits.
Over the past two decades, numerous researchers have dedicated their efforts to unveiling, detecting, mitigating, and preventing these bugs, with the last decade witnessing a surge in research within this domain. Among the spectrum of concurrency bugs, data races or race condition vulnerabilities stand out as the most prevalent, accounting for a staggering 80\% of all concurrency bugs.
This survey paper is focused on the realm of race condition bug detectors. We systematically categorize these detectors based on the diverse methodologies they employ. Additionally, we delve into the techniques and algorithms associated with race detection, tracing the evolution of this field over time. Furthermore, we shed light on the application of fuzzing techniques in the detection of race condition vulnerabilities.
By reviewing these detectors and their static analyses, we draw conclusions and outline potential future research directions, including enhancing accuracy, performance, applicability, and comprehensiveness in race condition vulnerability detection.


\end{abstract}

\section{Introduction}
\begin{figure}[h]
    \centering
    \includegraphics[scale=0.67]{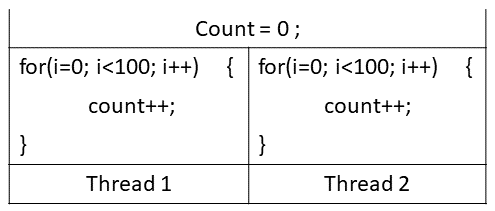}
    \caption{A simple depiction of data race condition}
    \label{fig:race1}
\end{figure}

The current era witnessed the emergence of Industry 5.0, also known as the Fifth Industrial Revolution, characterized by the collaboration between humans and advanced technology. This industrial paradigm shift involves the integration of humans and AI-powered machines, such as robots, to enhance workplace processes\cite{industry5}. As a result, industries like textiles, agriculture, healthcare, and transportation are leveraging information and communication technology to drive efficiency\cite{industry5.0}. Therefore, in the contemporary era of multi-core processors, harnessing concurrency has become a key driver for enhancing the performance of system software. As OS kernels and file systems evolve, an array of programming paradigms, including asynchronous work queues, read-copy-update (RCU), and optimistic locking, have been introduced to leverage multi-core computing capabilities \cite{lu2014study}\cite{min2016understanding}\cite{huang2016evolutionary}\cite{aghayev2019file}. However, these performance enhancements have not come without a cost. Concurrency bugs, particularly race conditions, have infiltrated the codebase, significantly compromising the reliability and security of file systems. These bugs can lead to deadlock situations, kernel panics, data inconsistencies, and privilege escalations \cite{mitre-cve-2009-1235}\cite{corbet2018unprivileged}\cite{fonseca2014ski}\cite{xu2019fuzzing}\cite{kim2019finding}\cite{kernel-btrfs-bugzilla}\cite{mitre-f2fs-cve}\cite{kernel-ext4-bugzilla}. Despite ongoing efforts to harden OS kernels against various attacks, such as kASLR, kCFI, and UniSan, these defenses primarily address memory errors (e.g., stack and buffer overflows) and have limited effectiveness in mitigating attacks that exploit concurrency bugs. Race conditions, in particular, represent a challenging class of concurrency bugs, wherein two threads erroneously access a shared memory location without proper synchronization or ordering, as depicted in Figure \ref{fig:race1}. Due to the inherent non-determinism in thread interleaving, race conditions are notoriously elusive, demanding precise timing and rare interleaving conditions to manifest. For instance, even widely used file systems like \textit{ext4}\cite{kernel-ext4-bugzilla} and \textit{ btrfs}\cite{kernel-btrfs-bugzilla}, with 50,000 and 130,000 lines of code respectively, witnessed an alarming number of reported bugs in a single year\cite{rodeh2013btrfs}. According to the Linux kernel bug tracker, there are currently 1,234 open bugs in Linux kernel 5.15.0. Of these, 627 are classified as critical, 307 are classified as serious, and 300 are classified as moderate.\cite{kernel-bugzilla}.
 In extreme cases, the failure to protect the system from concurrency vulnerability has led to a fatality. 
The Therac-25, a radiation therapy machine developed by Atomic Energy of Canada Limited (AECL), serves as a striking example of a significant threat posed by concurrent programming errors, commonly known as race conditions. Tragically, these errors occasionally led to patients receiving radiation doses hundreds of times higher than intended, resulting in fatalities or severe injuries\cite{therac}.

\subsection{Motivation}
\begin{figure}[ht]
    \centering
    \includegraphics[width=\linewidth]{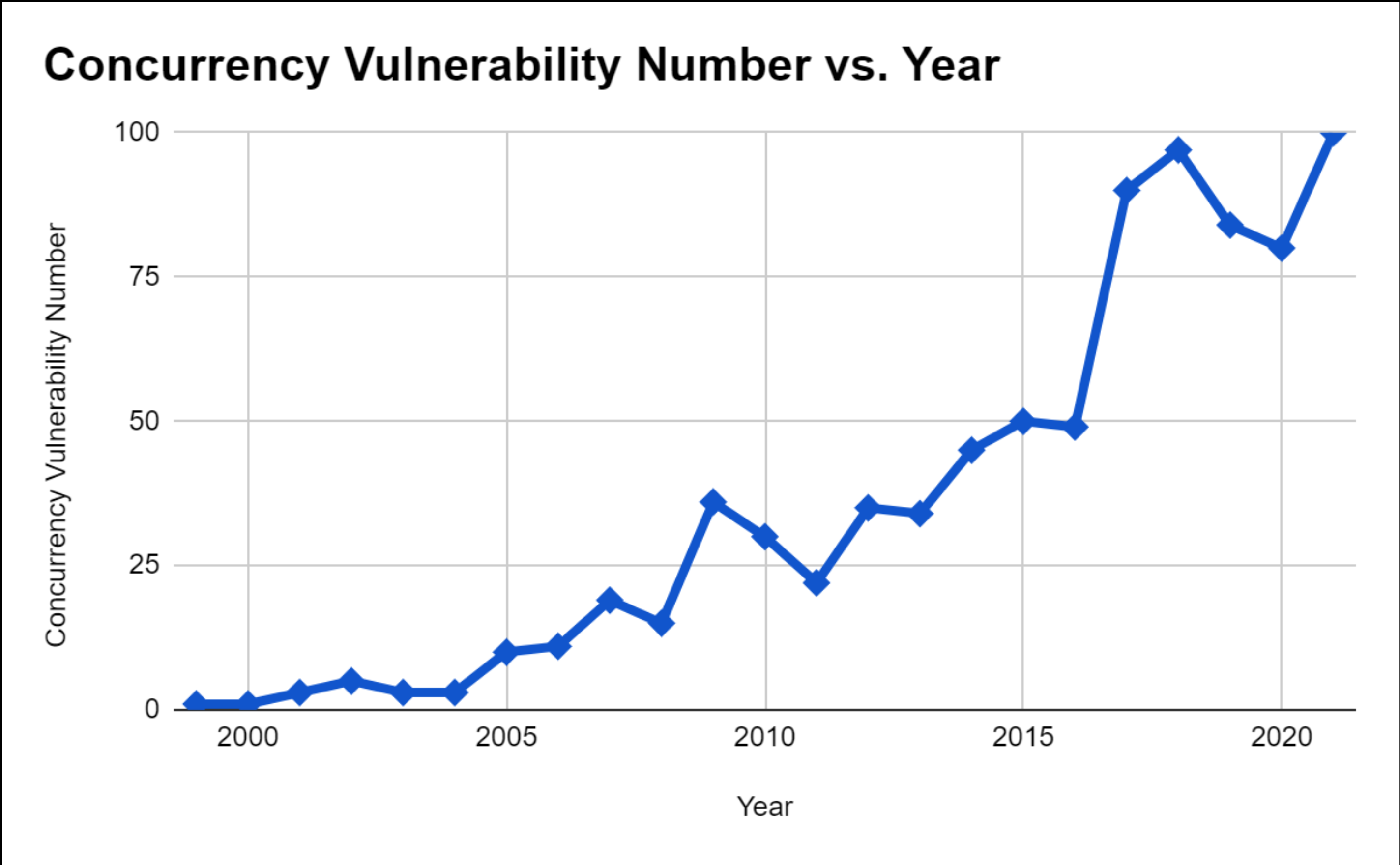}
    \caption{Increasing trend of concurrency vulnerability in last 20 years}
    \label{fig:trend}
\end{figure}

Recent hardware architectures have seen a significant increase in the prevalence of multithreading support, with processors like the Intel Xeon Platinum 9282 having up to 56 cores per socket, each capable of running 2 threads, and the Intel Xeon Phi Processor 7290F, an accelerator with up to 72 cores, each supporting 4 threads.Systems with six to 260 cores per socket were included in the 500 list of supercomputers.\cite{top500} Because these systems use simultaneous multithreading (SMT), many threads can be executed concurrently on operating systems that support SMT and/or Symmetric Multi-Processing (SMP).All threads in a node share the memory (not the cache), which allows for shared memory multithreading. The memory can be accessed uniformly or non-uniformly (UMA/NUMA). Language extensions or application programming interfaces (APIs) were once used by the scientific community to execute parallel programs written in languages like FORTRAN and C/C++ across several nodes in a cluster or grid. As multi-core computers became more common, the emphasis moved to shared memory programming models. This meant combining ordinary languages like C/C++ or specialized parallel languages like array-FORTRAN or HPF with tools like the pthreads library/runtime system.\cite{ieee2018}
However, the adoption of multiprocessing and parallel architectures requires careful consideration of potential bugs and challenges associated with concurrent software execution. Addressing these issues is crucial to harness the full potential of the rapidly evolving technological landscape.  The increasing prevalence of concurrent vulnerabilities and cyberattacks has underscored the need for secure IT infrastructure \cite{VUl}\cite{vul2}. The graph in Figure\ref{fig:trend} clearly shows how concurrency vulnerabilities have evolved from 1999 to 2021. Over this time frame, we see a consistent increase in the number of these vulnerabilities that have been reported\cite{CVEdetails}. This rise in vulnerability disclosures is closely tied to the growth of multi-core hardware and the increasing complexity of software that runs concurrently. It serves as a clear indicator of just how important it is to continue studying and researching concurrency vulnerabilities in the world of modern computing.
\\
Understanding the nuances of concurrent applications and the security implications they pose is critical \cite{li2017survey}. Researchers in the field have developed various tools and techniques to detect and mitigate vulnerabilities, particularly race conditions \cite{chen1990dert}. These vulnerabilities are prevalent in applications that utilize shared resources and often arise due to mistakes made by users and developers \cite{castro2001empirical}. Efforts to address these challenges have led to advancements in detecting and preventing race condition vulnerabilities \cite{yang2011finding}.

\subsection{Research Questions}
\begin{figure}[ht]
    \centering
    \includegraphics[width=\linewidth]{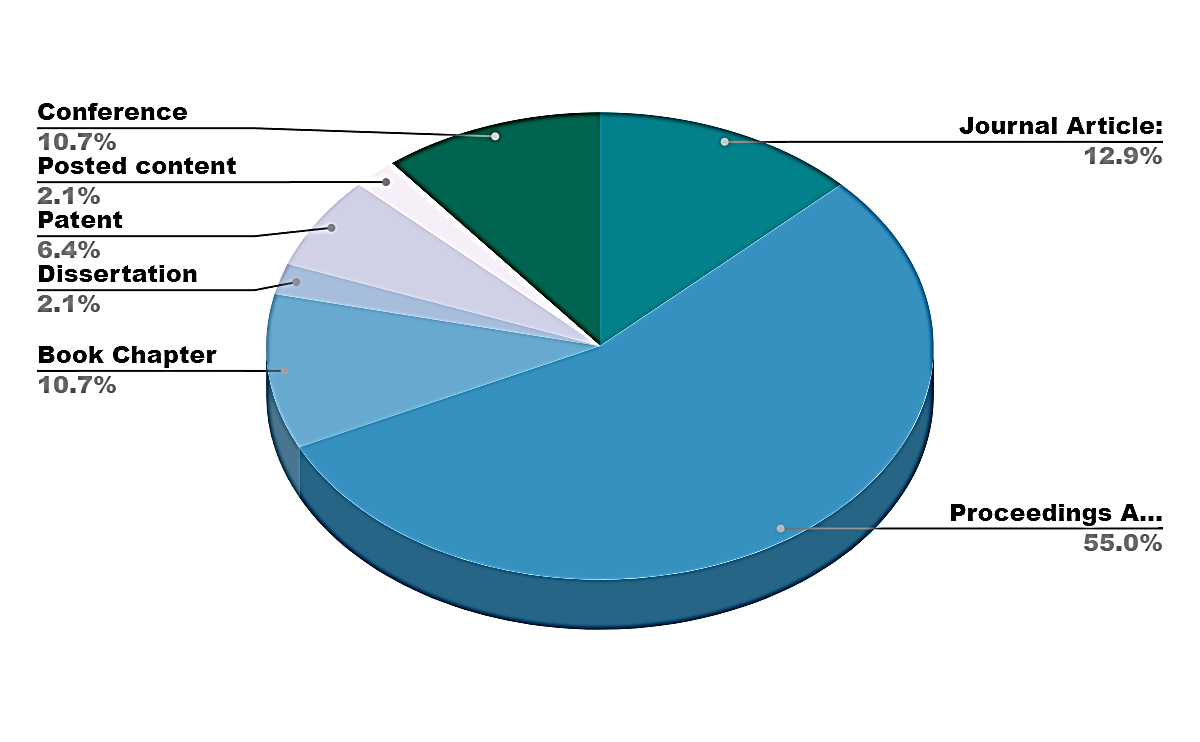}
    \caption{ Division of included papers}
    \label{fig:paperdis}
\end{figure}

This survey seeks to conduct a comprehensive and longitudinal analysis of race condition vulnerabilities in concurrent programming. Researchers in the field of multiprocessing and parallel computing have proposed various tools and techniques to detect, prevent, and mitigate these vulnerabilities. 
The review process involved a meticulous examination of available literature, achieved by systematically searching through journal articles and conference papers that tackled various aspects related to concurrency bugs. This encompassed methodologies, tools, techniques, empirical assessments, and surveys.
Major internet repositories were searched for pertinent papers. Initially, searches were conducted using the DBLP database, Google Scholar, and prestigious academic publishers including Wiley, ACM, IEEE, Springer, and Elsevier. Only those papers with certain keywords in their names, abstracts, or keywords—such as "concurrency bug," "concurrent program," "multithreaded program," "data race,"mutliprocessing"," "race condition," or "concurrency vulnerability"—were submitted.
In addition, a period of time spanning from 1993 to 2023 was set aside for the inclusion of articles, and those deemed to be of inconsequential importance or with defective ideas were purposefully excluded. Additional articles of noteworthy value that were not found in the previously listed databases were hand-picked and added during the evaluation process. 
 Figure \ref{fig:paperdis} shows the spread of the selected research and their publishing category. This paper surveys the concurrency bugs and the most up-to-date and well-known race conditions detectors. We categorize the existing detectors based on the types of techniques used to detect race condition vulnerability. Also, we shed special light on different types of fuzzers used in the detection of concurrency bugs. 
This paper aims to present an overview of the state-of-the-art in race condition vulnerability assessment and address the following research questions.
\\\textbf{RQ1: What causes concurrency vulnerabilities to arise and which type of concurrency bug makes up the majority of concurrency bugs}
\\ 
Before going to the detection methodologies of concurrent bugs we have made an in-depth study of the various concurrency bugs and the reason for their occurrence. We have also included in our study the importance of lock-free synchronization and the concurrency issues arising from them. 
\\ 
\textbf{RQ2: Study the different methods and techniques for the exposure and detection of data race bugs.}
\\
We have provided an overview of the current state-of-the-art race condition vulnerability assessment and analysis in hyper-threaded and concurrent applications. We have reviewed the methods and techniques that have been developed to identify these vulnerabilities, and we have discussed their strengths and limitations. We have created our own taxonomy of the techniques used by the detectors and on that basis classified them into three categories namely static, dynamic, and hybrid. Specifically, we have reviewed the fuzzers that have been developed to identify these vulnerabilities and discussed their strengths and limitations.
\\
\textbf{RQ3: What can be the future research aspect in the area of detection of race condition vulnerability?}
\\
Reviewing the most significant and current data race bug detectors which allowed to draw various conclusions about future research directions, including those related to accuracy, performance, application, and integrality.
\\
 We have also provided details of CWEs related to race condition vulnerabilities and described how they were discovered and remediated. Finally, we have outlined best practices that can be used to avoid or mitigate race condition vulnerabilities in hyperthreaded and concurrent applications.

\subsection{Review Organisation}
This literature review is structured into three primary sections to address research questions. In Section 2, we offer background information, including concepts and notation related to concurrent programming. Section 3 outlines the issues in concurrent programming, discussing the characteristics of each bug and its severity. In Section 4, we emphasize understanding race condition vulnerability, with a particular focus on data race bugs. Section 5 explores application analysis for vulnerability detection, examining three analysis techniques: static, dynamic, and hybrid analysis. Section 6 discusses tools and repositories for application analysis and vulnerability detection, including studies on code vulnerability detection using machine learning and traditional approaches. We categorize experimental studies into three core sections: application/code analysis, code vulnerability detection, and supporting tools. Section 7 addresses potential research directions, and we provide a concluding summary of our findings.

\section{Background}
Prior to conducting a security analysis, it's crucial to understand how multiprocessing systems operate. This section offers an overview of the evolution of concurrent applications, their operation, the significance of concurrent and hyperthreaded applications, and key technical concepts used in this paper. It also explores the use of multithreading and synchronization methods in concurrent applications, aiming to improve overall reader comprehension.
\subsection{Notations and Concepts of Concurrent Programming}
In the 1990s, C programming incorporated the POSIX thread library, known as pthreads, enabling multi-threaded software development. Today, modern computers and operating systems can run multiple programs concurrently, where concurrency means executing multiple instruction sequences simultaneously. This involves process threads that communicate through shared memory or message passing. However, concurrent resource sharing can lead to issues like resource scarcity and deadlocks. To improve efficiency, strategies like process coordination, memory allocation, and execution scheduling are employed. This section briefly introduces processes, threads, and computer architecture advancement, which are crucial for understanding vulnerabilities stemming from shared resources and concurrency.
\subsubsection{Processs and Threads}
\begin{figure}
    \centering
    \includegraphics[scale=0.55]{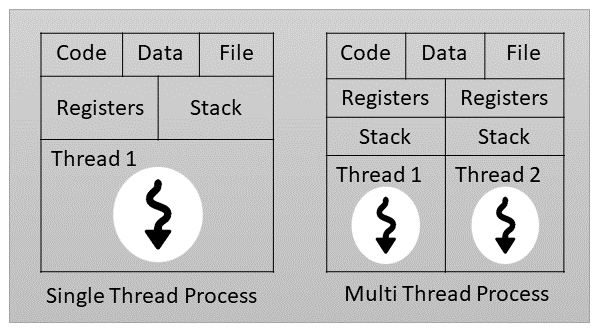}
    \caption{Multithreaded applications on a singel core}
    \label{fig:multithreading}
\end{figure}

Processes and threads are fundamental to concurrent computing, each serving distinct roles. Processes are encapsulated execution units that provide isolation by maintaining separate memory space and resources \cite{tanenbaum2014modern}. This isolation ensures robustness and fault tolerance in modern operating systems. In contrast, threads are lightweight execution units within a process, sharing the same memory space, which facilitates efficient communication and coordination among concurrent tasks \cite{butenhof1997programming}. This shared-memory model allows for parallelism exploitation, resulting in improved application performance and responsiveness.
The interaction between processes and threads is crucial in designing concurrent software systems. Processes offer high isolation levels, making them suitable for running independent tasks or safeguarding critical system resources \cite{silberschatz2018operating}. Threads, with their shared-memory model, enable efficient collaboration and parallelism within a process \cite{lewis1998threads}. Striking a balance between process and thread usage is a critical aspect of concurrent software design, ensuring resource utilization and efficient execution in modern computing environments.
Figure \ref{fig:multithreading} depicts the multithreading in a single processor systems. For a mutlicore or multiprocessor mutlithreading the system will have multiple copies of code , data and file with each core acting as a independent processor. 
\subsubsection{Synchronisation mechanism}
\begin{figure}[ht]
    \centering
    \includegraphics[width=\linewidth]{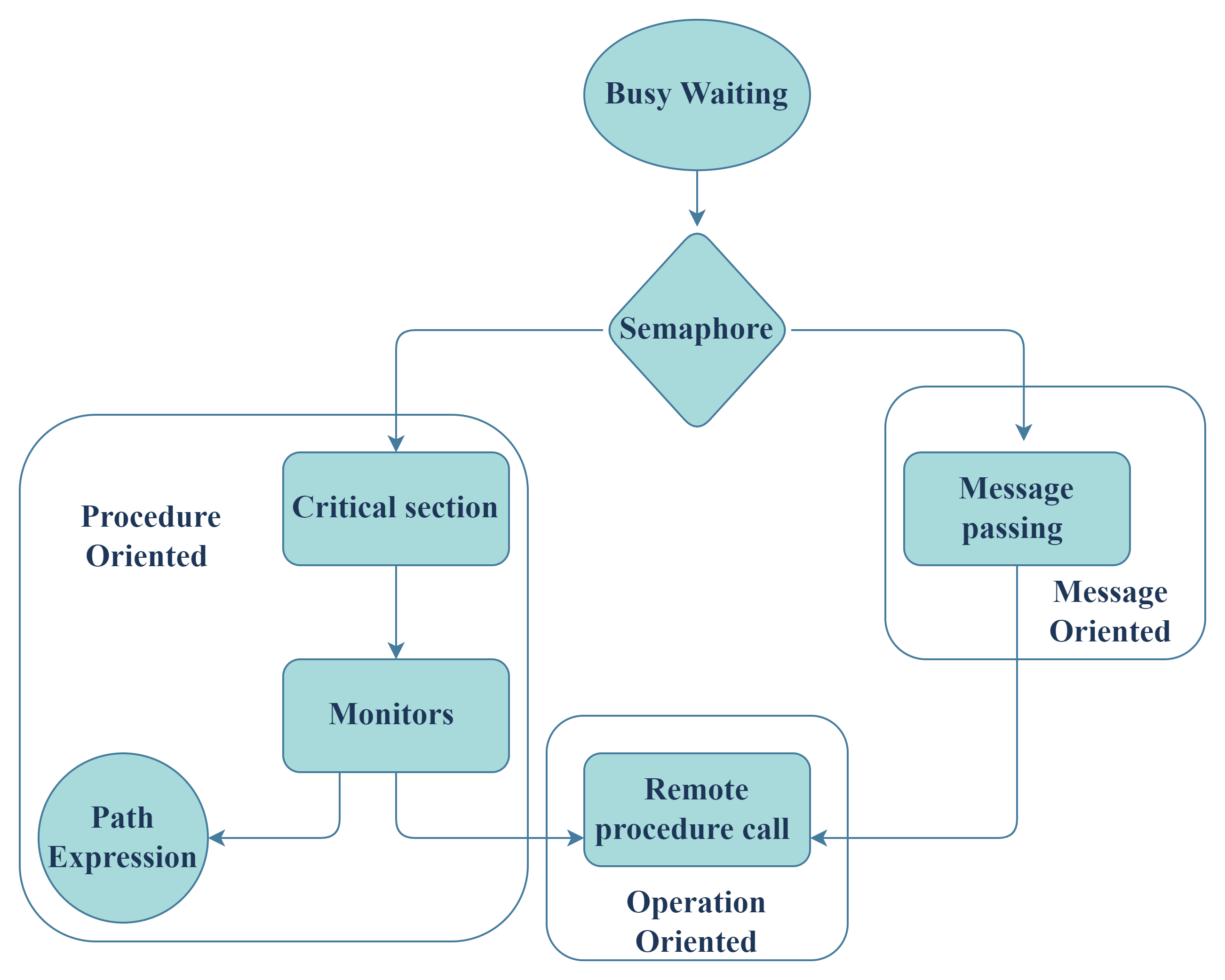}
    \caption{Synchronization techniques and language classes using locks.}
    \label{fig:sysnchronisation}
\end{figure}

OS employs sophisticated strategies to enable concurrent execution of multiple programs (processes) and concurrent sub-tasks within these processes (threads) \cite{silberschatz2014operating}. These mechanisms play a pivotal role in ensuring efficient system performance and enabling users to perform tasks simultaneously, contributing to a seamless and responsive computing experience.
\begin{itemize}
    \item \textbf{Locks}
    Mutex locks are commonly used to synchronize access to shared resources in multithreaded programs, allowing one thread at a time to access the shared resource. When a thread needs access, it must acquire the lock, use the resource, and then release the lock for other threads.
    Early work in operating systems identified two key synchronization types: mutual exclusion and condition synchronization, leading to the development of various synchronization primitives, such as busy-waiting semaphores, structured semaphores (e.g., critical regions, monitors, path expressions), message-passing primitives, and remote procedure calls. Busy-waiting semaphores, the earliest primitives, were found to be challenging and inefficient\cite{silberschatz2018operating}. Structured semaphores were introduced to address these issues, offering a more organized approach. Message-passing primitives enhance synchronization with data, and remote procedure calls combine procedural interfaces with message passing. This historical and conceptual relationship among these primitives is illustrated in Figure \ref{fig:sysnchronisation}.
    
    \item \textbf{Lockless synchronization}
Lockless synchronization, or lock-free synchronization, is a programming technique that manages shared resources without traditional locks or mutexes. It enhances the scalability and performance of multi-threaded or multi-core applications by reducing contention and minimizing blocking operations\cite{herlihy2008art}. This approach allows multiple threads to access shared resources concurrently, preventing data races and deadlocks. It's particularly useful in scenarios where locks would introduce significant overhead due to thread contention\cite{moir2007using}.
Lockless synchronization offers increased application parallelism, optimizing multi-core processor utilization. However, designing and implementing lockless algorithms can be intricate, necessitating deep knowledge of low-level memory operations and platform-specific intricacies\cite{harris2001pragmatic}.
The concept of Compare and Swap (CAS) is often integral to lockless synchronization, providing a foundation for atomic operations and ensuring thread safety. CAS allows threads to compare a value in memory to an expected value and update it if the comparison succeeds, all in a single, uninterruptible operation \cite{herlihy2008art}. This concept is crucial for implementing lock-free data structures and algorithms in lockless synchronization paradigms.
\end{itemize}

\begin{algorithm}
    \caption{Compare and Swap (CAS)}
    \begin{algorithmic}[1]
        \Procedure{CAS}{\textit{address}, \textit{expectedValue}, \textit{newValue}}
            \State \textit{currentValue} $\gets$ \textit{address}\;
            \If{\textit{currentValue} $=$ \textit{expectedValue}}
                \State \textit{address} $\gets$ \textit{newValue}\;
            \EndIf
            \State \textbf{return} \textit{currentValue}\;
        \EndProcedure {}
    \end{algorithmic}
\end{algorithm}

The Compare and Swap (CAS) algorithm is a vital tool in concurrent programming, providing a means to atomically update a shared variable while ensuring thread safety. CAS takes three inputs: the memory location \textbf{\textit{address}} to be updated, the expected value \textbf{\textit{expectedValue }}that the address should have, and the new value \textbf{\textit{newValue}} to set if the address matches the expected value. CAS reads the current value at the address, compares it to the expected value, and if they match, updates the address with the new value in an uninterruptible operation. CAS returns the current value regardless of whether the update occurs, making it useful for building lock-free data structures and synchronization primitives in multi-threaded applications, where only one thread at a time can successfully perform the update, ensuring coordinated and thread-safe modifications of shared variables.

\subsection{The Advancement of Computer Architecture}
\begin{figure}[ht]
    \centering
    \includegraphics[scale=0.5]{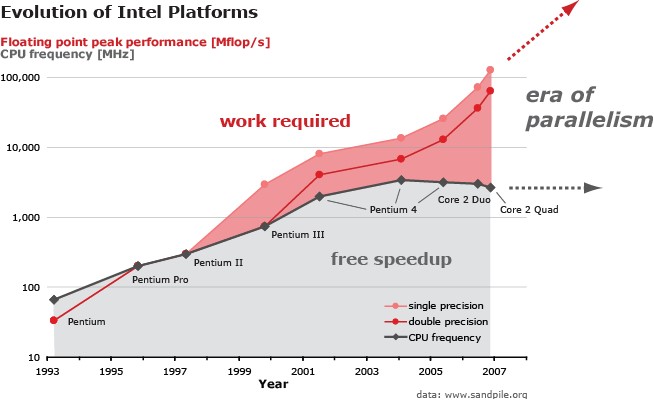}
    \caption{Evolution of Intel Platform over the years \cite{key}} 
    \label{fig:evolution}
\end{figure}

The concept of "Moore's Law" coined by Gordon Moore in 1965 predicted that the number of transistors on a single chip would double every year or two, resulting in increased computing speed\cite{Moore_intel}. However, as depicted in Figure \ref{fig:evolution}, the proliferation of transistors also led to overheating issues. This compelled a shift in technology towards multi-processing, driving significant advancements in computer architecture. These advancements aimed to create faster, more efficient, and highly parallel computing systems, fueled by factors like the demand for enhanced computational power, energy efficiency, and the emergence of new technologies\cite{Moore}.

One notable development is the transition from single-core to multi-core processors, enabling simultaneous execution of multiple tasks and enhancing system performance. Multi-core architectures have become commonplace in modern computing, facilitating parallel execution and improved throughput.\cite{hennessy2011computer}.
Another crucial advancement is the introduction of vector processing and SIMD (Single Instruction, Multiple Data) architectures. These technologies enable the simultaneous execution of a single instruction on multiple data elements, accelerating data-intensive computations. SIMD architectures find applications in scientific simulations, multimedia processing, and machine learning. \cite{hwu2011simd}.
Moreover, specialized accelerators like GPUs and FPGAs have revolutionized computer architecture. GPUs, originally designed for graphics, have evolved into powerful parallel processors for various computations. FPGAs provide hardware reconfigurability, allowing customization for specific workloads. \cite{li2017survey}.
Advances in memory technologies, including DDR4 and HBM, have alleviated memory bandwidth bottlenecks and facilitated efficient data access for high-performance computing.\cite{jacob2011memory}.
In summary, computer architecture has evolved to focus on parallelism and concurrency, as the era of increasing clock speeds has come to an end. Harnessing this computing power necessitates deep understanding and efficient software development.

\subsubsection{Multiprocessing Architecture}
\begin{figure*}[ht]
    \centering
    \includegraphics[scale=0.7]{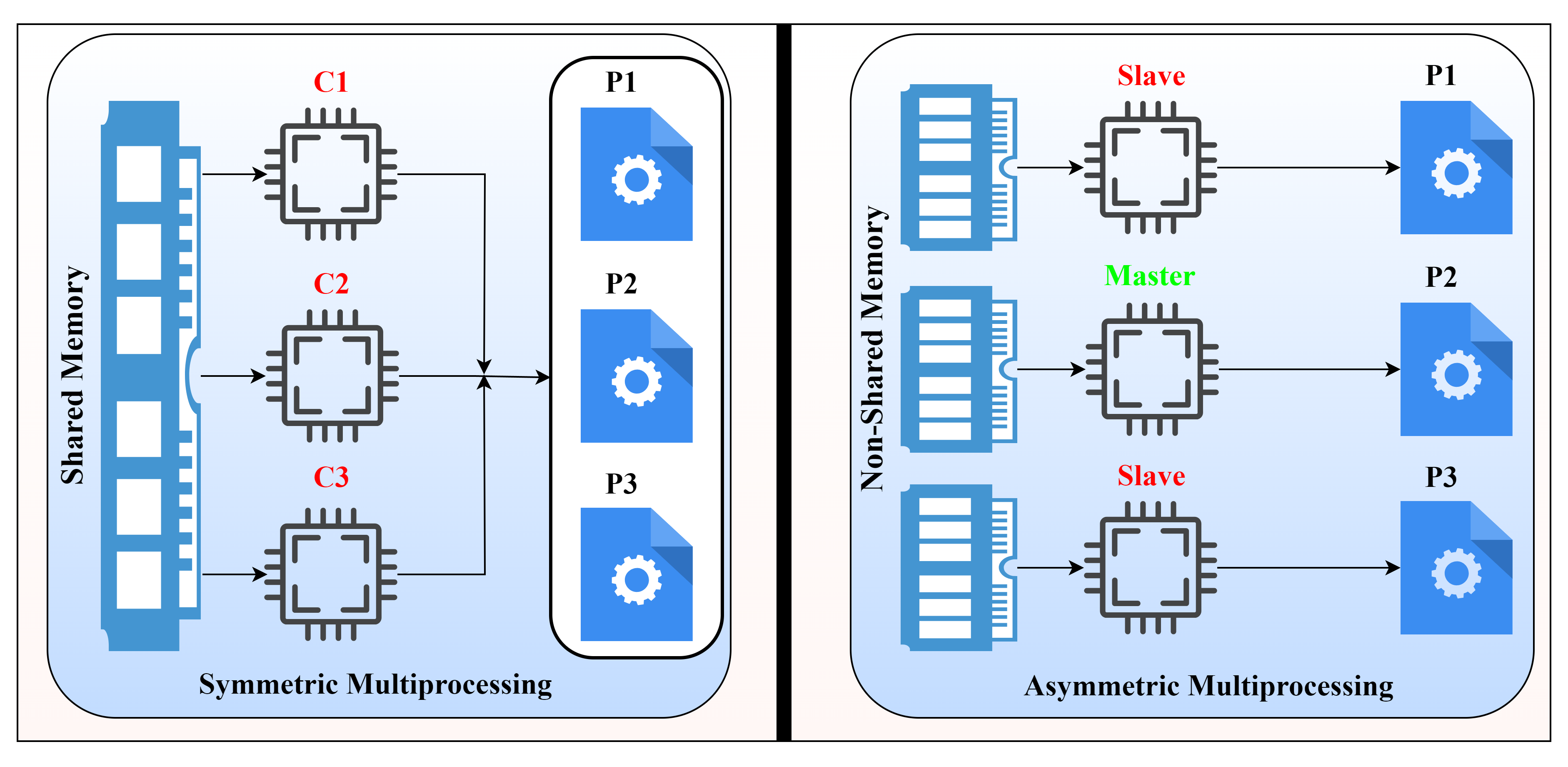}
    \caption{Symmetric and Asymmetric Multiprocessing}
    \label{fig:ASM}
\end{figure*}

Parallelism for program acceleration is a core objective of multiprocessor systems. Research in this area explores parallelism identification in sequential programs, resource allocation among competing processes, synchronization of cooperative processes, and the verification of parallel programs. The term "multiprocessing" often encompasses parallel processing systems, including multiprocessor systems.
There are two types of multiprocessing:
\begin{itemize}
\item \textbf{Symmetric Multiprocessing}
Symmetric multiprocessing (SMP) is a computing architecture where multiple processors or cores are identical in capabilities and roles. All processors share the same memory and operate under a single operating system, enabling parallel task execution and efficient resource utilization. SMP systems provide load balancing, scalability, and fault tolerance, making them suitable for various computing environments, including servers, workstations, and high-performance computing clusters. Challenges in SMP systems include memory consistency and synchronization, necessitating synchronization mechanisms like locks and semaphores to ensure data integrity \cite{dutta}\cite{bosch}\cite{silberschatz}\cite{dandamudi}.
    \item \textbf{Asymmetric Multiprocessing}
Asymmetric multiprocessing (AMP) is a computing architecture where multiple processors or cores have distinct capabilities and roles. This diversity allows for optimized resource allocation and enhanced system efficiency. Unlike symmetric multiprocessing (SMP), where processors have similar capabilities, AMP systems categorize processors into roles, such as application processors for general tasks and auxiliary processors for specific functions like graphics rendering or I/O operations. AMP is commonly used in mobile devices, where heterogeneous multi-core processors optimize power consumption and balance performance. It is also applied in high-performance computing (HPC) environments, leveraging processor capabilities for parallelism and specific workload acceleration. Effective task scheduling and load balancing are critical in AMP systems, given varying processor capabilities \cite{gupta}\cite{kumar}\cite{lebel}.
\end{itemize}
Figure\ref{fig:ASM} shows the underlying difference between symmetric and asymmetric multiprocessing systems, where the symmetric multiprocessing architecture works on a shared memory principle whereas the asymmetric multiprocessing system works in the master-slave model.

\subsubsection{Hyperthreading}
\begin{figure*}[ht]
    \centering
    \includegraphics[scale=0.5]{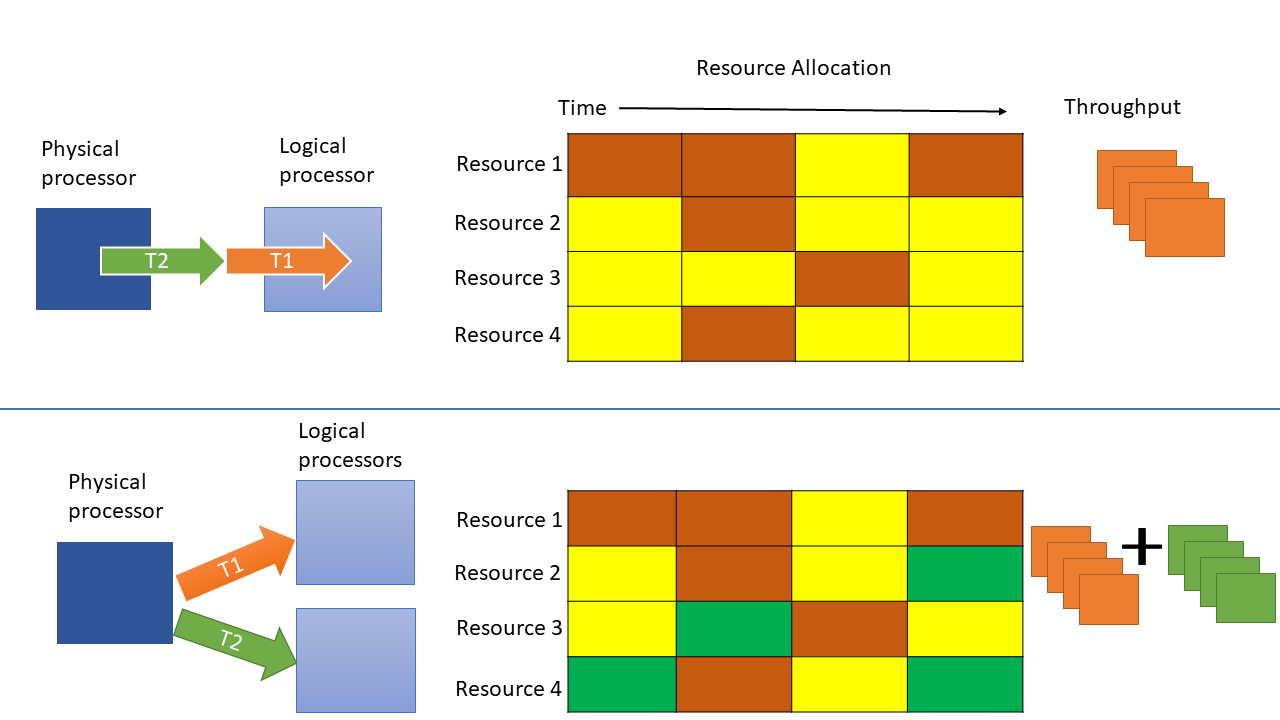}
    \caption{Performance Improvement using Hyperthreading Technology}
     \label{fig:hyperthreading}
\end{figure*}

Various multithreading strategies have been employed over the years to address memory latency issues. One approach is Simultaneous Multi-Threading (SMT), which exposes greater parallelism to the CPU by handling instructions from multiple streams, thus enhancing processor utilization. Notably, SMT, such as Intel's Hyper-Threading (HT), is a cost-effective design choice that does not require extensive additional hardware \cite{NASA}. Hyper-Threading, patented by Intel, is a technology present in Intel® XeonTM and Intel Pentium® 4 processors. It enables a single physical processor to appear as two logical processors to the operating system. By duplicating each processor's architectural state while sharing execution resources, it allows a single physical processor to execute instructions from multiple threads in parallel, potentially improving overall performance \cite{dell-ht}.
Figure\ref{fig:hyperthreading} gives the detailed working advantage of a hyperthreaded system over a single-core processor, where the throughput is increased due to hyperthreading technology.
Research has explored the performance impact of Hyper-Threading. NASA conducted an analysis of HT's effect on processor resource utilization in scientific applications, concluding that while HT enhances processor resource utilization efficiency, it does not consistently result in improved application performance \cite{NASA}. Another study found that Hyper-Threading on specific processors provided minimal performance gains and even degraded the effectiveness of certain applications, particularly those reliant on vectorization and facing communication bottlenecks \cite{HT-L1}.
Conversely, a study focusing on database management systems (DBMS) reported performance improvements with Hyper-Threading, specifically in TPC-C-equivalent and TPC-H-equivalent queries \cite{HTL2}. Additionally, research has explored the impact of Hyper-Threading on resource utilization, concluding that while it enhances processor resource utilization efficiency, it may not universally improve application performance. Efficient code optimization and leveraging vector units are vital for enhancing code efficiency and maximizing processor resource utilization, ultimately leading to improved application performance \cite{htl3}.

\section{Issues in Concurrency}
In the previous section, we examined the integration of multiprocessing systems and the advancements in hyperthreading technology, both of which have contributed to the widespread adoption of concurrent programming. This section will address the complexities and challenges associated with concurrent programming, along with an exploration of the distinct characteristics of concurrency-related software bugs.
As understood in previous sections the concurrent programs use different synchronization primitives to coordinate between the threads and how the program will execute. The failure of these synchronization mechanisms can lead to various bugs such as data races, deadlock, atomicity violation, live lock, ABA problems \& spurious wakeup.
 Broadly, concurrency-related bugs can be categorized into four major types: data races (race conditions), order violations, violations of atomicity, and deadlocks, we include the challenges related to the ABA problem (arriving from lockless synchronization) and spurious wake-ups as well in this section.
 \begin{table*}[ht]
 \caption{Concurrency vulnerability causes and mitigation }
 \label{tabel:concu bug}
 \begin{tabular}{p{4cm}p{6.5cm}p{6.5cm}}
 \hline
 \textbf{Concurrency Vulnerability} & \textbf{Causes} & \textbf{Mitigation} \\
 \hline
 Race Condition & Arises when several threads or processes use the same data at the same time and at least one of them writes something.& \checkmark Access to shared data can be coordinated by using synchronization techniques like locks, semaphores, or atomic actions. \\
& & \checkmark Employ thread-safe programming practices and avoid sharing mutable state between threads without proper synchronization. \\
 \hline
 Deadlock & Occurs when a circular dependency is created between two or more threads or processes that are blocked as they wait for resources that they jointly own.
 & \checkmark Employ resource allocation and release strategies to ensure proper resource management and prevent circular dependencies. \\
 & & \checkmark Use techniques like resource ordering or resource preemption to break potential deadlocks. \\
 \hline
 Starvation & Happens when a thread or process is consistently refused access to resources or execution because other threads or processes are given a higher priority.
 & \checkmark Implement fair scheduling policies to prevent indefinite starvation. \\
 & & \checkmark Employ strategies to prevent priority inversion, such as priority ceiling protocols or priority inheritance.\\
 \hline
 Data Inconsistency & Arises when multiple threads or processes perform concurrent read and write operations on shared data without proper synchronization, leading to inconsistent or incorrect results. & \checkmark Use atomic operations, locks, or transactions as synchronization techniques to guarantee correct consistency and coordination. \\
 & & \checkmark Apply proper data access patterns and avoid unsynchronized concurrent modifications.\\
\hline
ABA problem & Two or more threads modify a shared variable, leaving the variable with the same value as it had before the modifications. & \checkmark In order to guarantee that only one thread is able to change the shared variable at once, use a compare-and-swap (CAS) procedure.\\
 & & \checkmark Use a version stamp to track the number of times the shared variable has been modified.\\
\hline
Spurious wakeup & A thread is awakened from a wait state even though no other thread has signaled it. & \checkmark Use a wait-free or lock-free algorithm to avoid spurious wakeups.\\
 & & \checkmark Use a retry loop to repeatedly attempt the wait operation until it succeeds.\\
\hline
\end{tabular}
 \end{table*}
 
\subsection{Data Races}
Data races occur when multiple threads simultaneously access the same part of computer memory (like a shared piece of information) without the proper coordination, and at least one of them is trying to change it. This can cause memory operations to happen unexpectedly, leading to different threads seeing inconsistent data and values. Surprisingly, sometimes, programmers intentionally allow data races for performance reasons, although it's important to note that data races and race conditions, despite often being used interchangeably, refer to different things. Data races relate to improper coordination in memory access, while race conditions are about timing-dependent issues during execution.
\subsection{Atomicity Violation}
Atomicity violation is a type of problem in multithreading, and it accounts for nearly 70 percent of all reported issues in this category. It happens when one thread's actions disrupt the order of operations in another thread, particularly when both threads are using the same shared resource. This disruption can cause the program to behave unpredictably.
\subsection{Order Violation} 
Order violation is another issue after race conditions and atomicity violations. It arises due to the reordering of operations involving different parts of memory. In simpler terms, when two operations, let's call them A and B, are supposed to happen in a specific order (like A always before B), but during execution, that order gets mixed up. Fixing an order violation often leads to a deadlock, a situation where threads get stuck waiting for each other. Interestingly, order violation is the least studied among these issues, often confused with atomicity violation. Research shows that both programmers and testers often struggle to identify the correct sequence of thread execution, making it a challenging problem to solve. It's also noteworthy that fixing errors in multithreaded applications usually takes longer compared to single-threaded ones.
\subsection{Deadlock} Deadlock is an error in multithreaded programming when threads become stuck, unable to gain control over a mutex, which is vital for preventing issues like race conditions. Deadlocks typically arise from improper sequencing of operations or failure to release the mutex correctly. They can be classified into resource and communication deadlocks, with the former being more frequent. Four conditions must be met for a deadlock to occur: mutual exclusion (resources used by only one thread at a time), holding and waiting (a thread holds one resource and waits for another held by a different thread), no expropriation (only the holding thread can release a resource), and recurring wait (threads forming a cycle while waiting for each other's resources). All four conditions must be satisfied for a deadlock to manifest. Common scenarios for deadlock errors include mutexes mutually excluding each other, missing mutex release operations, attempting to create a mutex in a loop, and repeated calls to a function that generates a mutex.
\subsection{ABA Problem}
The ABA problem is a concurrency bug that occurs in lock-free and wait-free algorithms. It happens when one thread reads a shared variable, another thread alters the variable's value and then reverts it to the original value before the first thread writes to it. This can result in concurrency issues as the first thread remains unaware of the variable's change, leading to incorrect assumptions, unexpected behavior, or crashes. For instance, in a lock status scenario, if one thread acquires the lock (setting the variable to 1), and another thread waits for it to return to 0, an ABA problem can arise. If the first thread reacquires the lock, the variable changes back to 1, leaving the second thread blocked indefinitely. Detecting and resolving the ABA problem can be complex due to intricate interactions between threads, but mitigation techniques like using compare-and-swap (CAS) operations or version numbers can help prevent it.
\subsection{Spurious Wakeup}
\begin{figure}[h]
    \centering
    \includegraphics[width=\linewidth]{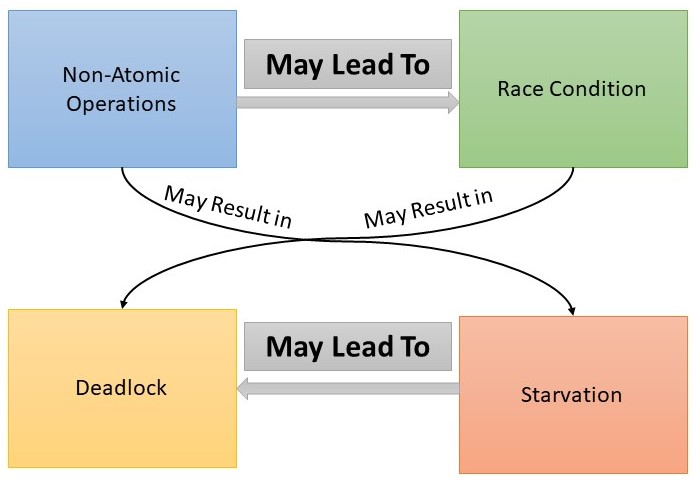}
    \caption{Relationship between the concurrency bugs}
    \label{fig:relation}
\end{figure}

A spurious wakeup is like an unexpected event in a multi-tasking program. It happens when a part of the program wakes up for no clear reason, usually when it's waiting for something to happen. This unexpected wakeup can lead to problems like strange behavior or even crashes in the program. For example, think of a program that uses a special signal to control when different parts can use a shared resource. If a part of the program wakes up without getting the right signal, it might try to use the shared resource at the wrong time, causing problems like data errors and other unexpected issues. Detecting and fixing spurious wakeups can be tricky because they often happen in complex situations with many parts of the program working together. However, there are ways to make spurious wakeups less likely, like using signals that can handle these unexpected events and using special tools to protect shared data.
\\
A summarisation of the concurrency bug with their possible causes and mitigation is explained in Table\ref{tabel:concu bug}. From the definitions above and the Figure\ref{fig:relation} as well we can infer that all the concurrency vulnerabilities are interrelated and there is a likelihood that a program containing a concurrency bug is very likely to trigger atmost one other concurrency bug.
In their extensive 2022 study, Lilibo et al. meticulously analyzed 839 concurrency vulnerabilities sourced from the National Vulnerability Database (NVD), revealing crucial insights into this prevalent issue. These vulnerabilities have surged in response to the growing scale and complexity of software systems, posing significant security threats. The most common among them are race conditions, particularly notorious for their potential to lead to severe remote attacks across diverse services. These concurrency vulnerabilities primarily fall into the medium-risk category, making up 67.2\% of cases, although the distribution is relatively even among high and low-risk instances. Notably, both local and remote network access are used for exploitation, with many attackers favoring concurrency vulnerabilities due to their ease of exploitation. The study identified \textbf{\textit{race conditions}} as the primary culprits across varying severity levels, often resulting in denial of service issues. Furthermore, these vulnerabilities can lead to system file leaks, notably affecting system availability, with a higher risk of complete loss after exploitation\cite{lilibo}. 

\section{Race-Condition Vulnerability}
\begin{figure}[h]
    \centering
    \includegraphics[width=\linewidth]{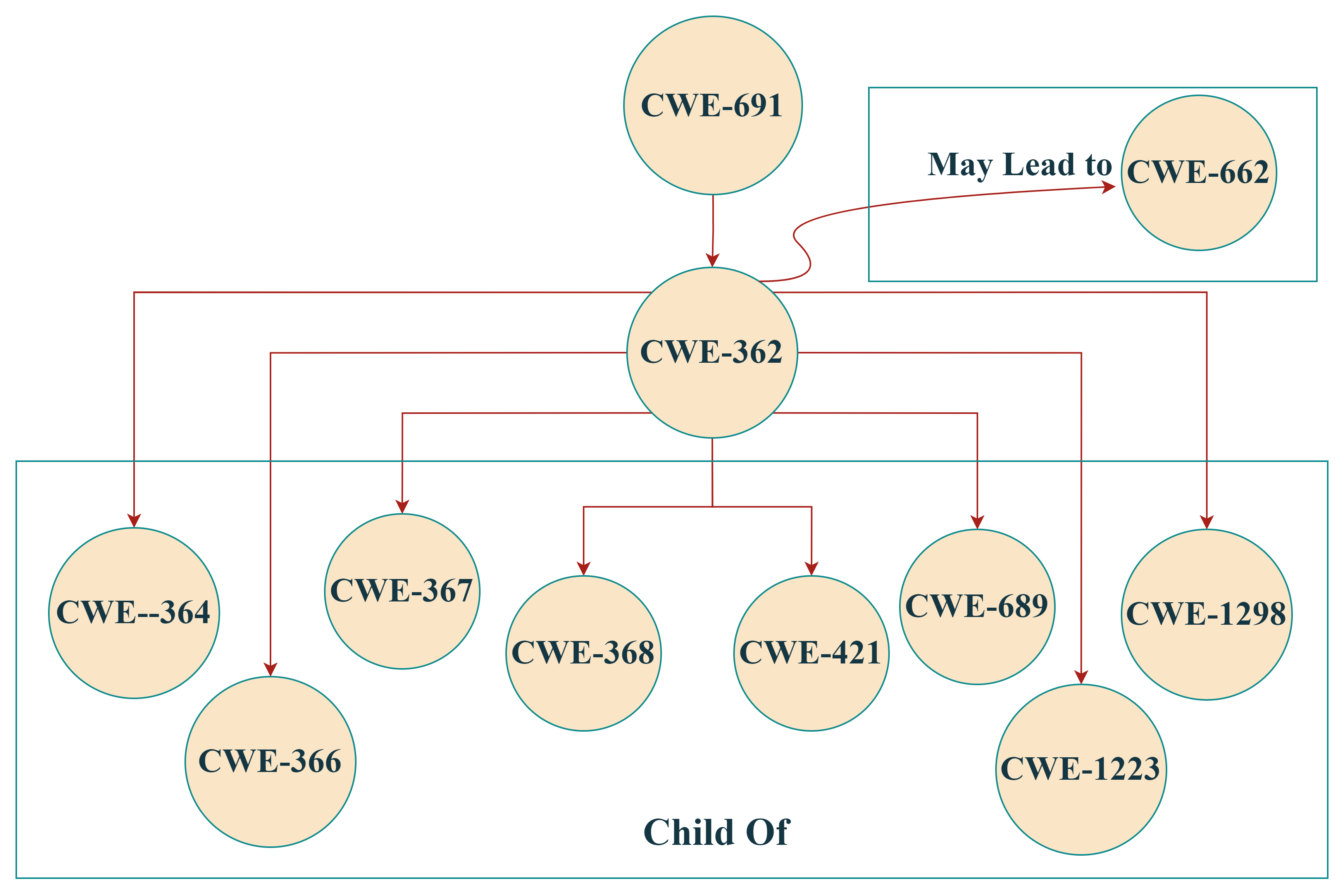}
    \caption{Relationship chart for CWE-362}
    \label{fig:relationship cwe}
\end{figure}

The analysis of Lilibo et al. study\cite{lilibo} reveals the critical role of race conditions in security concerns. therefore, in this and the preceding sections, we discuss everything keeping in view race condition vulnerability.
In this section, our primary focus is on race condition vulnerabilities, a type of software flaw that arises when multiple execution contexts, such as threads or processes, share a shared resource and can modify it concurrently, disregarding the need for mutual exclusion. An incorrect assumption often made is that a sequence of instructions or system calls will execute atomically, preventing any interference from other threads or processes. Regrettably, some developers tend to downplay the significance of addressing this issue, even when presented with evidence. In reality, most system calls involve the execution of thousands, or even millions, of instructions and often don't complete until another process or thread is given a chance to run.
\\
The race condition vulnerability can occur in different parts of a computer system including : 
\begin{itemize}
    \item \textbf{Multithreaded Applications: }When multiple threads in a program access shared resources concurrently and perform read and write operations without proper synchronization, race conditions can arise.
    \item \textbf{Operating Systems:} When many processes or threads access shared kernel resources, like system data structures or device drivers, without the proper synchronization, race conditions can arise within the kernel or operating system components.
    \item \textbf{Network Communication:} Race conditions can also occur in network communication protocols when multiple processes or systems try to access and modify shared network resources concurrently.
    \item \textbf{Database Systems: }When several transactions try to access and modify the same data concurrently in database systems without the necessary separation mechanisms—such as locks or transaction management—race situations may occur.
    \item \textbf{File Systems: }Concurrent access to shared files or directories by multiple processes or threads can lead to race conditions if proper file locking or synchronization mechanisms are not employed.
\end{itemize}
All the above-listed sections where race condition can occur in a computer system can be mapped with the different CWEs related to race condition as depicted in Figure \ref{fig:relationship cwe}. Based on data association and operation block, Hong et al. have illustrated four types of race bugs in \cite{datarace}: data race bug, block race bug, multi-data race bug, and multi-data block race bug.
 All these four classes can be related to the CWEs listed in Table\ref{table:relate_table}. Figure\ref{fig:relationship cwe} gives a relational structure of CWE-362 "Concurrent Execution using Shared Resource with Improper Synchronization" also known as race condition. An improper synchronization of shared resources is a cause of multiple CWEs which are said to be the child of the CWE-362. 
\begin{table*}[!ht]
    \centering
    \caption{CWEs Related to Race Condition Vulnerabilities in Various Applications/Components}
    \label{table:relate_table}
    \begin{tabular} {p{5cm}p{1.5cm}p{10cm}}
    \hline
        \textbf{Name } & \textbf{CWE -ID } & \textbf{Description }  \\ \hline
        Insufficient Control Flow Management  & 691 & The code lacks adequate control flow management during execution, resulting in potential scenarios where the control flow can be unexpectedly altered. \\ \hline
        Concurrent Execution using Shared Resource with Improper Synchronization ('Race Condition') & 362 & The product incorporates a code sequence that can execute simultaneously with other code. This code sequence necessitates temporary, exclusive control over a shared resource. However, there is a specific time interval during which the shared resource may be altered by another concurrently running code sequence.  \\ \hline
        Signal Handler Race Condition & 364 & The product utilises a signal handler that creates a race condition. \\ \hline
        Race Condition within a Thread & 366 & When two threads of execution concurrently access a resource, there is a potential risk of using the resource in an invalid state, which can result in an undefined state of execution. \\ \hline
        Time-of-check Time-of-use (TOCTOU) Race Condition & 367 & The product performs a pre-use verification of a resource's state. However, it is possible for the resource's state to be altered after the verification and before its utilisation, thereby rendering the verification results invalid. The occurrence of this issue may result in the product executing actions that are not valid, particularly when the resource is in an unforeseen or abnormal state.  \\ \hline
        Context Switching Race Condition & 368 & The product executes a sequence of non-atomic operations to transition between contexts that span privilege or security boundaries. However, a race condition exists, which enables an attacker to manipulate or distort the product's behavior during the transition. \\ \hline
        Race Condition During Access to Alternate Channel & 421 & The product provides an additional communication channel for authorized users, however, it is important to note that this channel may also be accessible to other actors.  \\ \hline
        Permission Race Condition During Resource Copy & 689 & The product does not establish the permissions or access control for a resource during the process of copying or cloning. As a result, the resource remains vulnerable to external entities until the copying process is finished.  \\ \hline
        Race Condition for Write-Once Attributes & 1223 & The hardware design includes a write-once register that can be programmed by an untrusted software component before the trusted software component. This sequence of events leads to the occurrence of a race condition issue. \\ \hline
        Hardware Logic Contains Race Conditions & 1298 & The presence of a race condition within the hardware logic has the potential to compromise the security guarantees provided by the system. \\ \hline
        Improper Synchronization & 662 & The product employs multiple threads or processes to enable temporary access to a shared resource that can only be accessed exclusively by one process at a time. However, it lacks proper synchronization of these actions, potentially resulting in simultaneous access to this resource by multiple threads or processes. \\ \hline
    \end{tabular}
\end{table*}
\\

To understand better, here is an example of a code snippet that intentionally contains a race condition:

\begin{lstlisting}[language=Octave]
def race_condition(x):
    # Acquire a lock on the variable x.
    lock.acquire()

    # Modify the variable x.
    x = x + 1

    # Release the lock on the variable x.
    lock.release()

# Create a lock object.
lock = threading.Lock()

# Create two threads that will both call the race_condition() function.
thread1 = threading.Thread(target=race_condition, args=(1,))
thread2 = threading.Thread(target=race_condition, args=(2,))

# Start the threads.
thread1.start()
thread2.start()

# Wait for the threads to finish.
thread1.join()
thread2.join()

# Print the value of x.
print(x)
\end{lstlisting}

The value of x will be printed to the console using this code. Nevertheless, if we execute this procedure more than once, we might observe various values for x. This is a result of the two threads simultaneously changing the variable x. When this happens, there may be a race condition and x does not end up as expected.
We can use a lock to make sure that only one thread can access the variable x at a time in order to avoid race situations. Before changing the variable x in the code above, we first constructed a lock object and obtained the lock. We changed the variable x and then released the lock. This guarantees that the variable x cannot be modified simultaneously by the two threads.
\\
\subsection{Data Race Bugs}
\begin{figure*}[h]
    \centering
    \includegraphics[width=\linewidth]{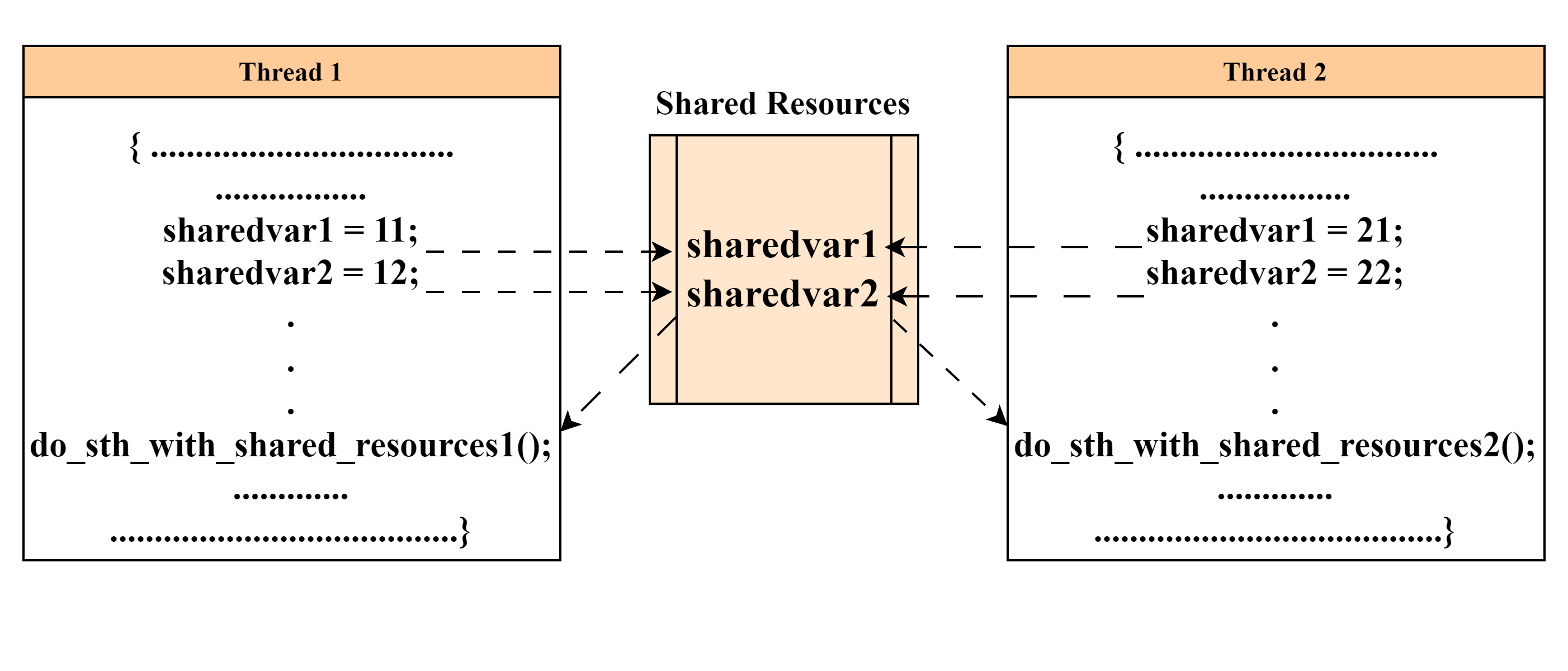}
    \caption{Simultaneous access to shared resources by two tasks without specific protection which can lead to data race}
    \label{fig:datarace}
\end{figure*}

When two or more threads visit the same shared variable concurrently and at least one of those accesses is a write, it might lead to a specific kind of race condition known as data race. Data corruption and deadlocks are two examples of unexpected or improper behavior that can result from data races. This survey's primary goal is to evaluate the approaches used to identify data races, both solely and inclusively.
When we look at the elucidation of data race though Figure\ref{fig:datarace} where two threads, Thread 1 and Thread2, update the values of shared resources, namely sharedvar1 and sharedvar2. Later on, these tasks retrieve the values of these shared resources using the functions $do\_sth\_with\_shared\_resources1()$ and $do\_sth\_with\_shared\_resources2()$, without any protective measures in place for these operations.
In the absence of protective measures, it is important to note that the expected value of a shared resource, such as sharedvar1, may not necessarily match the value that was written in Thread 1 just before the function call. This discrepancy arises due to the concurrent execution of Thread 1 and Thread 2, which can lead to unexpected outcomes, such as reading a value of 21 or, in some cases, even a corrupt, random value.
To elucidate this, consider the two possible sequences of events
Sequence 1Thread 1 writes sharedvar1 as 11 $\rightarrow$ 
Thread 1 invokes the do\_sth\_with\_shared\_resources1() function.
Sequence 2:Thread 1 writes sharedvar1 as 11 $\rightarrow$ Thread 2 overwrites sharedvar1 with a value of 21 $\rightarrow$ Thread 1 subsequently calls the do\_sth\_with\_shared\_resources1() function.
\\
In the absence of protective mechanisms, any code implemented within the do\_sth\_with\_shared\_resources1() function cannot assume a specific sequence of events or a predetermined value for sharedvar1.
Therefore, if your code relies on a particular value of sharedvar1, it can lead to unintended consequences, rendering the data race a software bug. Data races manifest when shared resources are accessed concurrently by multiple tasks, resulting in unpredictable outcomes. Understanding data races can be challenging since the execution of instructions may not follow the order in which they are written, and the outcome can vary between different test runs, making data races elusive and difficult to reproduce and rectify.
These findings emphasize the critical importance of implementing protective mechanisms to detect and mitigate data races and ensure the reliability and integrity of concurrent software systems. Therefore our proceeding sections will concentrate on the \textbf{various detection techniques and tools for data race bugs}.

\section{Analysis for Vulnerability}
The Program under test(PUT) should be analyzed for vulnerability as a first step to the detection of vulnerability. Based on the analysis techniques used by the researchers of all the papers that were considered for review, the papers were categorized into three basic detection techniques namely static, dynamic, and hybrid. Figure \ref{fig:techniquesl} depicts the division of the various techniques into these three categories. The three categories can be defined as :
\begin{figure}[h]
    \centering
    \includegraphics[scale=0.78]{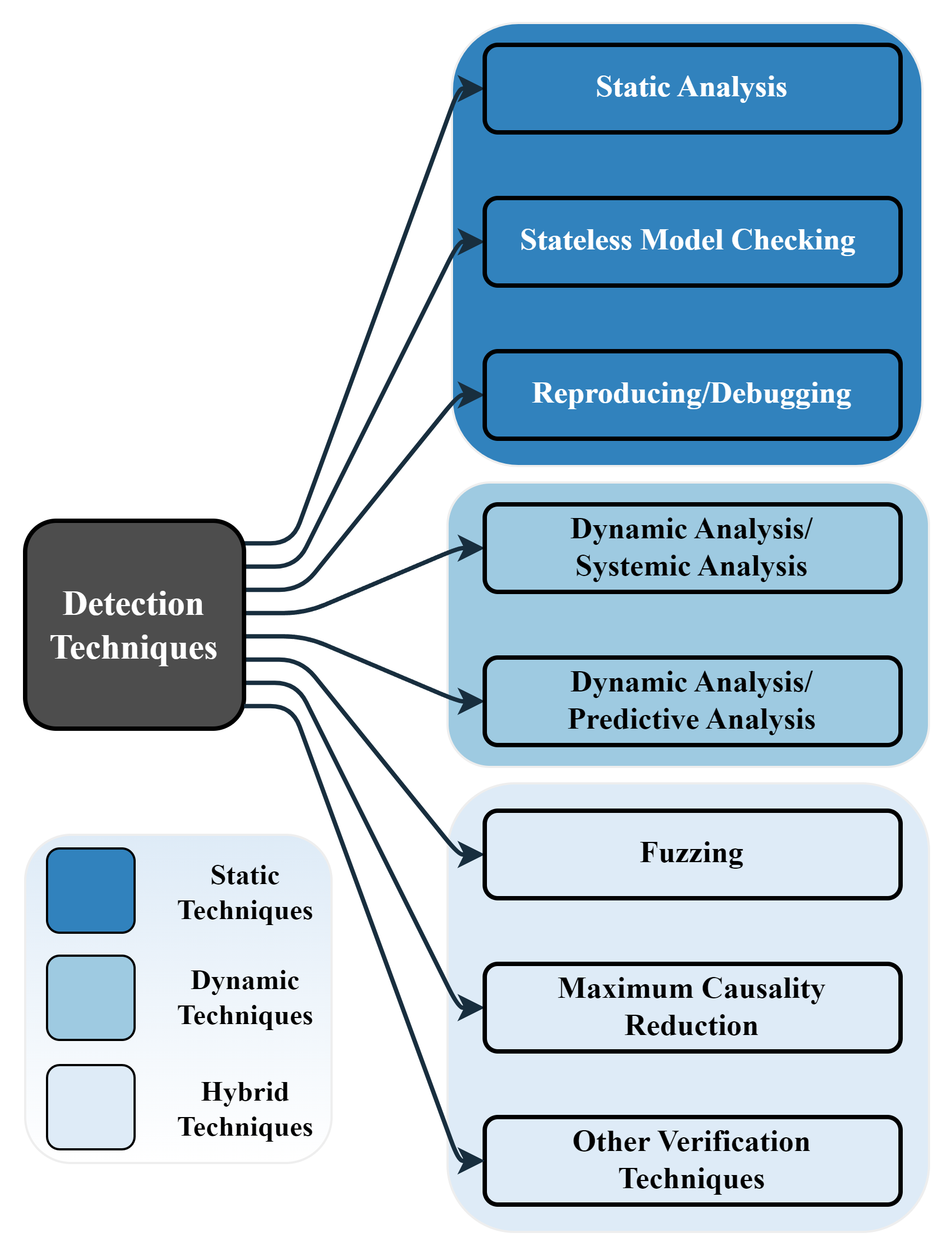}
    \caption{Categorisation of different detection techniques used in the reviewed papers}
    \label{fig:techniquesl}
\end{figure}
    \begin{enumerate}
    \item Static Analysis:Potential vulnerabilities can be found before the application is even compiled by using static analysis, which examines an application's code without running it.

    \item Dynamic Analysis: In dynamic analysis, the program is run and its behavior is observed for indicators of vulnerability, including race situations or deadlocks.

    \item Hybrid Analysis: The benefits of both static and dynamic are combined in hybrid analysis. Static analysis is usually the first step, in which the code is examined for possible problems before running. Certain vulnerabilities can be found and problems can be detected early in the development phase thanks to static analysis. Next, the software is run and the behavior is observed under real-world situations during the dynamic analysis phase. Runtime-specific bugs, performance difficulties, and security vulnerabilities that might not be visible with static analysis alone can be found during this phase.

\end{enumerate}

\subsection{Algorithms Detection of Race condition}
 Detection  algorithms are used to identify races in concurrent programming. To identify these races, numerous algorithms have been created, each with a different strategy and set of trade-offs. The following are some popular techniques (which include lock and lockless synchronization) for race detection:

\subsubsection{The lockset algorithm} 
This approach, which is used in both static and dynamic analysis tools, detects a possible race problem when several threads access shared memory without any of them having a shared lock. Essentially, the technique determines that for every shared memory variable, v, there is a non-empty set of locks, C(v), that each thread that wants to access the variable has to hold. At first, all of the locks that are available are placed in 'C(v)'.
Every thread has two sets of locks: 'writeLocks(t)' indicates the write locks held, and 'locks(t)' indicates all the locks held by that thread. The following is how the algorithm functions:

\begin{enumerate}
    \item For every shared memory variable 'v,' initialize 'C(v)' with the complete list of locks.
    \item Each thread keeps track of its held locks ('locks(t)') and its write locks ('writeLocks(t)') during execution.
    \item Whenever a thread accesses a shared memory variable 'v':
        \begin{itemize}
            \item It checks whether it holds all the locks in 'C(v)' (i.e., 'locks(t)' contains all the locks in 'C(v)').
            \item It also verifies whether it holds the write lock for 'v' (i.e., 'writeLocks(t)' contains 'v').
        \end{itemize}
    \item If both conditions are met for a thread when accessing 'v,' there is no potential race condition detected. However, if any of the conditions are not satisfied, it signifies a potential race condition, as shared memory 'v' is being accessed without the necessary locks.
\end{enumerate}
Static and dynamic analysis tools employ this approach to identify possible race situations when many threads access shared memory without the necessary locks. 
It maintains sets of locks for each shared variable, ensuring that threads adhere to the locking protocol when accessing shared memory to prevent data races.
Regrettably, not all of the races identified by a lockset algorithm represent real races. It is possible to write code that is free from data races, either by employing clever programming techniques or by utilizing alternative synchronization methods such as signaling. This poses a significant challenge when it comes to distinguishing genuine bugs from false positives. Annotations and specific suppressions can be employed as strategies to mitigate this issue.
An alternative algorithm for race detection is the "happens-before" algorithm.

\subsubsection{Happen Before Algorithm}
The Happens-Before algorithm is designed to determine the partial ordering of events in distributed systems, particularly in the context of identifying data races. Here's an overview of how it operates
A single thread's events are naturally arranged according to the order in which they occur.
Events are arranged among threads based on synchronization fundamental features. In the event that two threads are vying for the same lock ('lock(a)'), for example, the unlocking of one thread is seen as having occurred before the locking of another thread.
A possible race issue is indicated when many threads visit the same variable and their accesses are not deterministically ordered by the "happens-before" relationship.

\begin{figure}[h]
    \centering
    \includegraphics{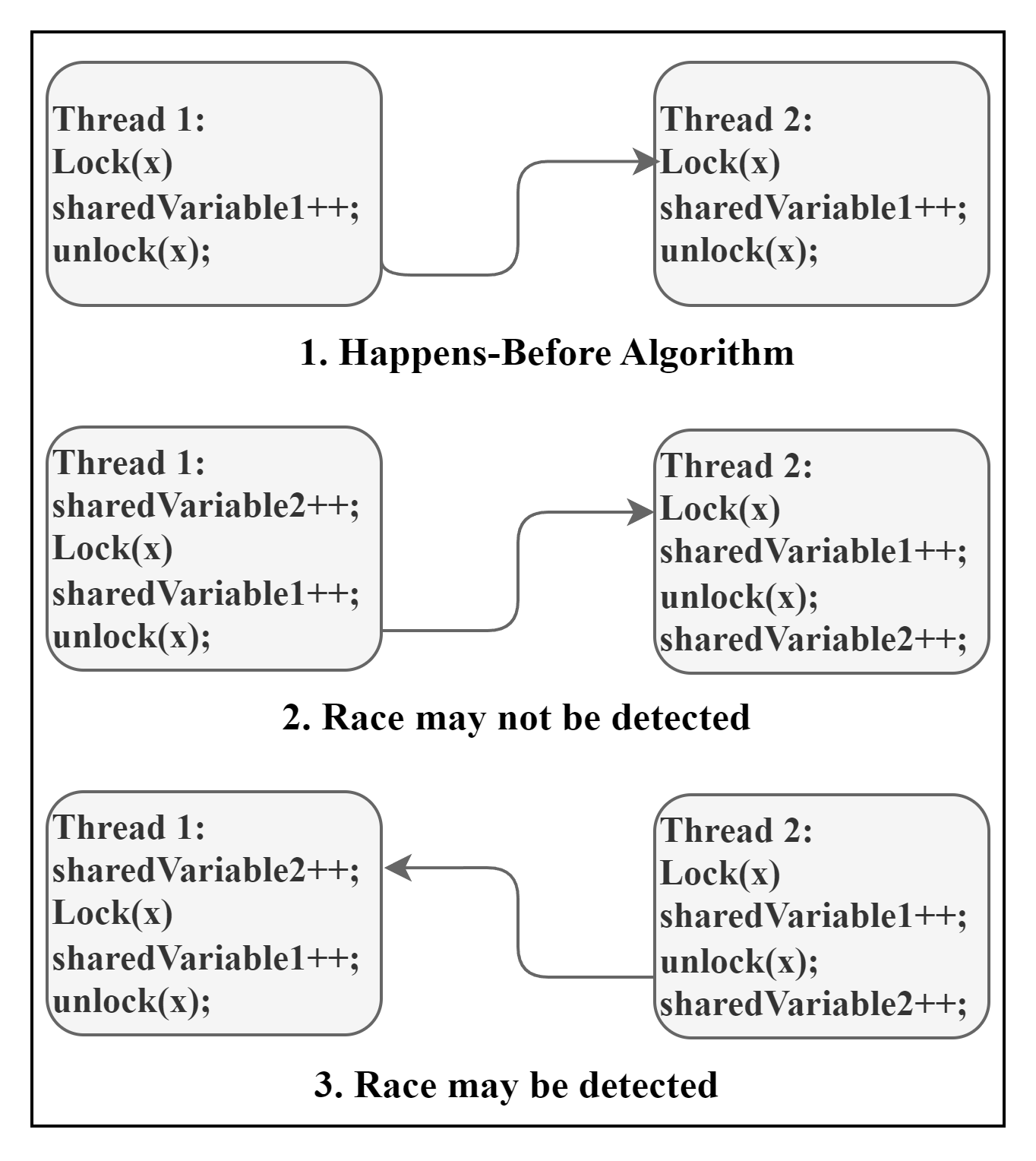}
    \caption{Happen Before Algorithm}
    \label{fig:happen before}
\end{figure}
However, there are some limitations to this algorithm:
It can be computationally expensive to monitor these relationships.
The order in which events are scheduled has a significant impact on this algorithm's performance. The partial order constructed is specific to a particular scheduling instance and may not detect the same bugs if the scheduling is different on another day.
While some executions (Fig\ref{fig:happen before}(2)) may not report any races, others (Figure\ref{fig:happen before}(3)) may detect races. This inconsistency can lead to missed race conditions, with some only becoming apparent years after a product's release.

Comparably, although the Lockset algorithm is effective, it frequently produces a large number of false positives. Attempts have been made to integrate these algorithms in order to take advantage of their individual advantages and minimize their disadvantages.
It's worth noting that race detection algorithms have undergone evolution over several years with the usage of a combination of other techniques such as\textit{ point analysis, CFG(Control flow graphs), reachability analysis and lock analysis} etc. 
\subsection{Fuzzing}
Fuzzing is an approach that begins by identifying possible concurrency issues using bug detectors. It then takes control of thread scheduling and execution based on these bug reports, aiming to execute specific interleaving patterns in the program's operation to uncover concurrency bugs intentionally. To extract traits that indicate the existence of concurrency flaws in real-world scenarios, several fuzzing approaches rely on static analysis. When fuzzing is done intentionally, it exposes concurrency flaws more effectively than other techniques. Fuzzing falls under the hybrid analysis category since it employs both static and dynamic techniques to find concurrency issues.

\subsection{Machine learning}
Machine learning or specific deep learning methods are also applied by researchers for the detection of concurrency bugs. There are several methods that deep learning can be applied to identify concurrency issues in concurrent programs. Using a dataset of acknowledged concurrency bugs to train a deep learning model is one such method.
The model can then be used to identify new concurrency bugs in other programs by comparing them to the known bugs in the dataset. Another approach is to use deep learning to generate synthetic execution traces of concurrent programs and then train a model to identify concurrency bugs in the traces. This approach can be more effective than traditional static analysis tools because it can detect concurrency bugs that are only exposed during runtime. Deep learning can also be used to develop dynamic analysis tools that can detect concurrency bugs in real-time. These tools can be used to monitor concurrent programs while they are running and to identify concurrency bugs as they occur.
DeepRace\cite{deeprace} is one such method that leverages deep learning techniques to automatically detect data races, eliminating the need for manual creation of data race detectors. To find data races in code approaches, it uses a convolutional neural network (CNN) that has one layer and several window sizes. To extract the weights of the final convolutional layer, it also integrates the class activation map function with global average pooling. The lines of code linked to data races are then identified by backpropagating these weights across the input source code. As a result, it is claimed that the DeepRace model may identify data race issues in both files and lines of code.

\section{Tools and Techniques for Detection}
\begin{figure}[ht]
    \centering
    \includegraphics[scale=0.5]{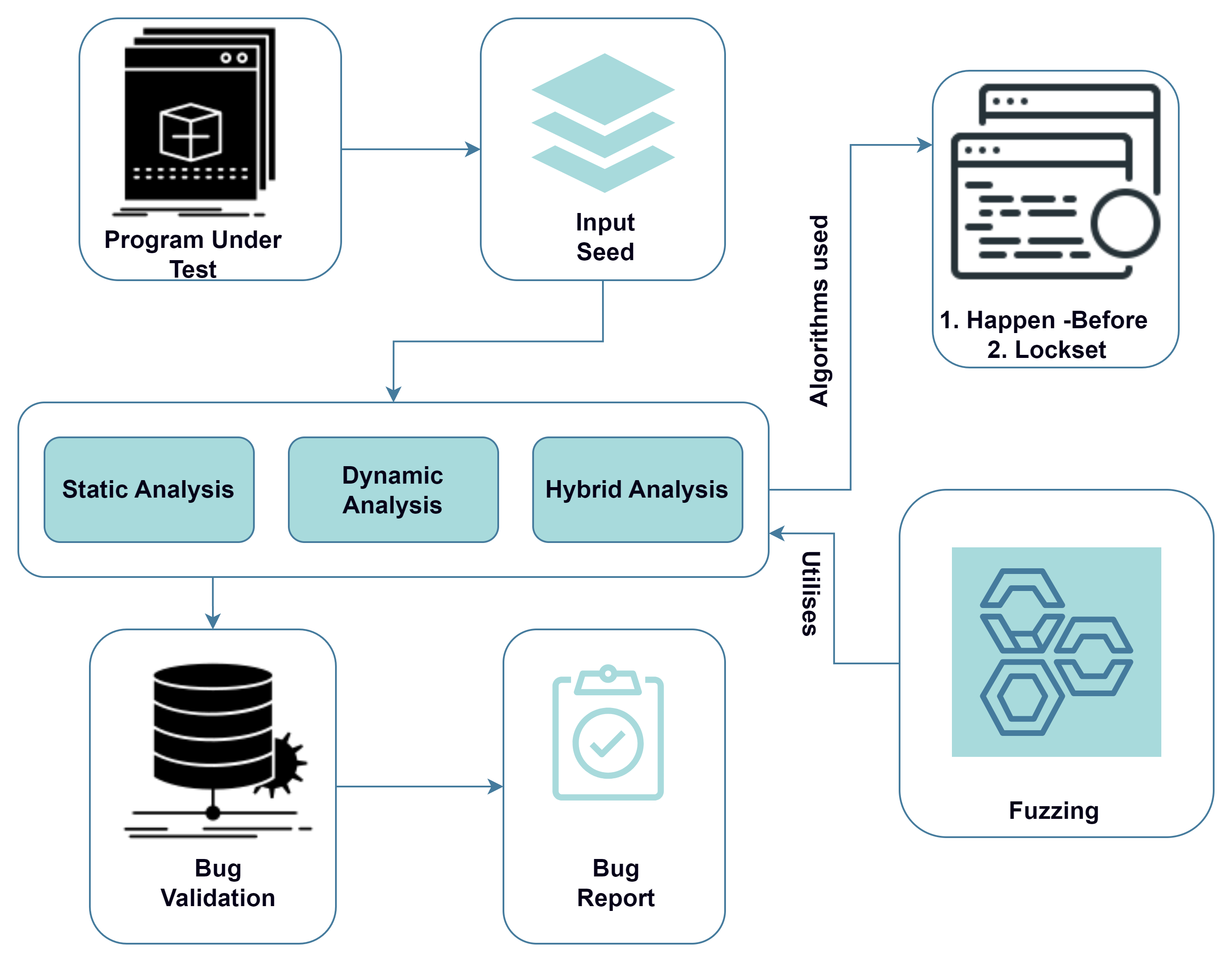}
    \caption{Flow of Concurrency Detection}
     \label{fig:flow}
\end{figure}

To Detecting concurrent application vulnerabilities can be challenging, as they often involve complex interactions between multiple threads or processes. Figure\ref{fig:flow} illustrates the comprehensive process of identifying concurrency bugs in a concurrent program (PUT) using an input seed. The various methods for detecting these concurrency bugs can be categorized into three distinct groups based on whether they involve program execution. These categories are static analysis, dynamic analysis, and hybrid analysis.
\\

In handling potential race conditions, hangs, and all the other problems that arise when processing simultaneous transactions, a variety of concurrency testing tools are available on the market. Every tool we examine here is useful in a specific capacity.

\subsection{Static Analysis Tools}
Because concurrency issues in software are intimately related to the way the program's scheduler manages threads, they are difficult to find using standard compile-time checks and testing techniques.
To address this, various static analysis techniques have been developed.
A static race detection analysis for concurrent Java programs was introduced by Flanagan and Freund et al.\cite{flanagan2010threadsafe}. Their system focuses on lock-based synchronization rules for avoiding races. It checks to see if locks are appropriately held while using shared variables. But using this approach necessitates adding more type annotations, which might add to the complexity.
To address this, Abadi and colleagues presented a static race detection study for Java programs running in high quantities concurrently\cite{abadi2006types}. There are no restrictions on this type-based analysis concerning test coverage issues. It works well with classes that have client-side or internal synchronization, as well as thread-local synchronization. Using 4000 lines of Java code, they tested this technique and were able to find race problems in other test programs as well as common Java libraries.
Engler presented RacerX\cite{racerx}, a program that can identify deadlocks and race conditions without the need for annotations. Facebook created the static analysis tool RacerX. By examining the code, keeping track of shared variable accesses, and looking for order violations, it finds race situations in Java programs.RacerX uses flow-sensitive analysis and concurrent inter-procedural program analysis to efficiently find problems in complex systems. Large applications can be handled by it, but its false-negative rate is somewhat high.
Naik et al. suggested a novel static method for Java program race detection in order to improve RacerX. Their four stage algorithm (reachable pairs, aliasing pairs, escaping pairs, and unlocked pairs) is context-sensitive. Using context sensitivity, it increases scalability even in large, extensively used programs and finds more flaws than earlier static race detection methods, such as RacerX.
Naik and Aiken et al. proposed an algorithm to overcome RacerX's shortcomings in demonstrating racial freedom. Their main goal is to show that if two locks are different, then the memory regions they guard have to be different as well. In order to statically analyze multi-threaded programs and identify races, they use conditional must-not-aliasing.
In order to counter code-centric techniques, Vaziri et al. presented a data-centric approach\cite{vaziri}. As a new criterion for data races, they identified 11 suspicious interleaving patterns and statically examined the code parts that required alteration to stop data races.
Voung et al. introduced RELAY\cite{relay}, a static and scalable method for identifying data races in the Linux kernel, in the context of C programs. In contrast to RacerX, RELAY detects data races through a context-sensitive, bottom-up analysis. In order to examine function behavior without regard to the calling context, it presents the idea of relative locksets. In 4.5 million lines of C code, RELAY detected 53 races; however, its false-positive rate was more than 70
To identify concurrency in Java programs, several tools have been developed \cite{coverity}\cite{threadsafe}.A commercial static analysis tool with race situation detection features is called Coverity\cite{coverity}. It tracks variable accesses and thread synchronization actions to perform inter-procedural analysis and find possible data races.
A static analysis tool called ThreadSafe \cite{threadsafe} is intended to find concurrency problems in Java programs, including race conditions. In order to find potential data races, deadlocks, and thread-safety violations, it examines the code.
A commercial static analysis tool with race condition detection capabilities is Polyspace \cite{polyspace}. Formal verification and abstract interpretation approaches are employed to identify possible concurrency-related problems, such as data races, in C/C++ systems.On the other hand, CogniCrypt \cite{cognicrypt} is a static analysis tool designed to find data races in cryptographic code in Java systems. It finds possible cryptography-related race situations by combining symbolic execution and program slicing.
These methods encompass various approaches to static and dynamic analysis for detecting concurrency bugs in both Java and C programs, each with its strengths and limitations.
Table\ref{table:static} depicts the comparative analysis of all the multiple techniques used by the static analysis tools for data race detection. Of all the tools majority of them use Happen Before and Point Analysis algorithms for the detection methodology. 
The use of CFG(Control flow graph) is also seen for the visualization of the program flow which makes it easier to statically analyze the pieces of code where the bug can occur. But, as the complexity increases  static analysis cannot be scaled and can lead to many false positives therefore there arises a need for dynamic and hybrid analysis
\begin{table*}[]
\centering
    \caption{Comparative Analysis Of Techniques Used by Static Detectors}
\label{table:static}
{%
\begin{tabular}{|p{1.9cm}|p
{2cm}|p{1.2cm}p{1.2cm}p{1.2cm}p{1cm}p{1cm}p{1.2cm}|p{4cm}|}
\hline
 \multirow{2}{*}{\textbf{Name of Tool}} &
\multirow{2}{*}{\textbf{Intended Use}} &
  
  \multicolumn{7}{c|}{\textbf{Methodology  Employed}} \\ 
  \cline{3-9}
   &
   &
   
 \textbf{Happen Before Analysis }& 
  \textbf{Point to Analysis} &
 \textbf{Reachability Analysis} &
  \textbf{Model Checker} &
 \textbf{DAG/CFG} &
 \textbf{Lock Set Analysis} & 
  \textbf{Description} \\ 
  \hline

RacerX \cite{racerx} &
  Linux, Free BSD &
 
  \checkmark &
   &
   &
  \checkmark &
   \checkmark &
   \checkmark  &
  Flow sensitive, interprocedural analysis \\
\hline

RELAY\cite{relay} &
  Linux Kernel & 
 \checkmark &
   &
   &
  \checkmark &
  \checkmark &
  \checkmark &
  Relative Lockset \\
\hline

ERIGONE\cite{erigone} & C language & \checkmark & & & \checkmark & \checkmark & \checkmark &  
  A partial reimplementation of spin model checker \\
\hline

RaceView\cite{Raceview} &
  Clanguage &
   &
   &
   &
   &
  \checkmark  &
   &
  DAG-based data race visualization for investigation and classification.\\
\hline

DR-Frame\cite{DRFRAME} &
  Network Applications &
  \checkmark  &
   &
   &
   &
 \checkmark  &
   &
  Based on dataflow analysis technique.Uses happen before algorithm. \\
\hline

SWORD\cite{sword} &
  Java Program written in eclipse IDE 
  &
\checkmark &
  \checkmark  &
   &
    &
   &
  &
 Points-to and happens-before analysis fusion for efficient and precise static data race detection. \\
\hline

Concurrent CFG\cite{concurrentcfg} &
  C language &
  \checkmark  &
 \checkmark  &
  &
   &
   &
   &
Points-to and happens-before analysis for static data race detection. \\
\hline

Chord\cite{chord}  &
  Java Program &
  &
 \checkmark  &
 \checkmark  &
   &
  &
 \checkmark &
  The proposed technique is based on a combination of points-to analysis, lock analysis, and reachability analysis \\
  \hline
  
LOCKSMITH\cite{locksmith} &
  C language &
   &
 \checkmark  &
   &
   &
  \checkmark &
 \checkmark &
 It uses a combination of interprocedural analysis and constraint solving to detect race conditions \\
\hline 

KISS\cite{kiss} &
Device Driver   &
  \checkmark  &
   &
   &
 \checkmark  &
   &
  \checkmark  &
Sequential data race detection on a simplified concurrent program \\
\hline

EPAJ\cite{EPAJ} &
Program Wirtten in JAVA &

  &
   &
   &
    &
  \checkmark  &
   \checkmark&
   Partial discovered types for static data race detection and atomicity checking with runtime performance optimization. \\
\hline

COBE\cite{cobe} &
   Program with asynchronous call &
  \checkmark  &
  &
   &
  \checkmark &
  \checkmark &
  \checkmark &
  Happens-before inference for asynchronous data race detection in concurrent programs.\\
\hline

IteRace\cite{iteracer} &
  JAVA Program &
  \checkmark  &
  \checkmark  &
  \checkmark  &
   &
  &
&
 Points-to, happens-before, and reachability analysis for precise and efficient static data race detection. \\
\hline

coderrect\cite{coderrect} &
  C/C++, JAVA /ANDROID APP &
 
 &
 \checkmark &
  &
 &
 \checkmark &
  \checkmark &
 Origin-based static race detection for precise and efficient reasoning about shared memory and pointer aliases. \\
  \hline
  
Open Race\cite{openrace} &
  C/C++ &
  \checkmark  &
  \checkmark  &
  \checkmark  &
 &
   &
 &
  Points-to, happens-before, and reachability analysis for sound and complete static data race detection. \\
  \hline
  
 LLOV\cite{LLOV} &
OpenMP &

  \checkmark  &
  \checkmark  &
  \checkmark  &
  &
  \checkmark  &
 &
  OpenMP control flow analysis-based data race detection for OpenMP programs using points-to, happens-before, and reachability analysis. \\
\hline 

OMPRacer\cite{ompracer} &
  OpenMP &
  
  \checkmark  &
   &
  \checkmark  &
  &
  \checkmark  &
 &
  OpenMP control flow analysis to construct a CFG of the OpenMP program and identify shared objects, followed by happens-before analysis and reachability analysis for data race detection. \\
\hline

Goblint\cite{GOBLINT} &
  Device Driver (Written in C,C++) &
  
  \checkmark  &
  \checkmark  &
  \checkmark  &
 &
  &
&
  Goblint uses points-to, happens-before, and reachability analysis for data race detection in device drivers.\\
  \hline
\end{tabular}%
}
\end{table*}

\subsection{Dynamic Analysis Tools}
Several dynamic analysis tools have been developed to detect data races and concurrency issues in various programming
languages. These dynamic methods offer certain advantages,
primarily the ability to examine observed feasible execution
paths and accurately assess variable values and thread interleaving modes. This flexibility allows dynamic methods
to effectively detect a range of issues, including plagiarism
detection, making them versatile in problem identification.
One notable tool in this category is Eraser\cite{eraser}, introduced
by Savage et al. in 1997. Eraser employs a dynamic approach
using the lockset algorithm to detect data races in multithreaded production servers. It ensures that shared-memory accesses adhere to programming policies, safeguarding programs
from data races. However, Eraser may have limitations in
error detection due to incomplete test coverage. Smaragdakis
et al.introduced the concept of “causally precedes” (CP), a
generalization of the happens-before algorithm, enabling the
observation of more races while maintaining accuracy and
completeness. This dynamic technique enhances the detection
of concurrency issues\cite{smaragdakis2012sound} Choi et al. presented an approach
that combines elements of both lockset and happens-before
algorithms to dynamically detect data races in object-oriented
programs\cite{choi2002efficient} Their experimental results demonstrated significant improvements in detection efficiency and reduced
overhead compared to existing techniques.
Yu et al. presented RaceTrack \cite{yu2005racetrack}, a useful runtime race detection tool for object-oriented applications, in opposition to Eraser and Choi's method. RaceTrack, which was created especially for Microsoft's Common Language Runtime, functions at the virtual machine level. It uses a hybrid detection approach to improve accuracy, reports questionable memory access patterns, and tracks program execution traces using instrumentation data. Xu et al. created the Serializability Violation Detector (SVD), a dynamic tool that offers root causes for debugging and employs backward error recovery (BER) to safeguard erroneous concurrent programs from errors\cite{xu2005serializability}. With SVD, one can examine something after the fact without having to first annotate the program. 
In order to identify error messages, Flanagan et al. underlined the significance of thorough and accurate dynamic analysis for atomicity violations\cite{flanagan2010threadsafe}. They did this by examining precise relationships between memory accesses in certain code areas.Ratanaworabhan et al.'s ToleRace\cite{ratanaworabhan2009detecting} addressed asymmetric races and used a transaction-like technique to detect and tolerate races, greatly lowering the overhead associated with dynamic race detection.
Jin et al. invented Cooperative Crug Isolation (CCI), a technique that may detect a wide range of concurrency flaws with minimal overhead and scalability, in contrast to many previous tools that target certain types of concurrency bugs\cite{jin2010instrumentation}. 
Falcon\cite{park2010falcon} by Park et al. 2010 was introduced as a dynamic fault localization tool that directly corresponds to faults, detects various types of concurrency bugs, and effectively captures both order violations and atomicity violations using patternbased analysis.
UNICORN \cite{park2012unified} enhanced Falcon by monitoring pairs of
memory access in C++ programs, allowing detection of both
single-variable and multi-variable violations.
A non-pattern-based tool called Recon\cite{lucia2011isolating} from Lucia et al. is used to handle single- and multivariable mistakes. It offers small segments of failure-inducing execution schedules to help with bug comprehension in addition to problem identification.
Zhang et al. introduced Anticipating Invariant (AI) as a
program invariant to detect and tolerate various concurrency
bugs, exposing order violations and generating emergency
patches when necessary\cite{zhang2016lightweight}
Some tools, like MagicFuzzer \cite{cai2012magicfuzzer} by Cai et al. 2012
and Magiclock \cite{cai2014magiclock} from Cai and Chan 2014, were developed exclusively for deadlock detection, identifying hidden
deadlock cycles and efficiently reducing the overhead of
dynamic deadlock detection. ConMem \cite{zhang2010conmem}  focused on bugs that can lead to program crashes caused
by incorrect thread interleavings and memory problems.To efficiently identify such problems, it keeps an eye on how programs are being executed and examines memory accesses and synchronizations.
ThreadSanitizer \cite{threadsanitizer}, developed by Google, is used for C/C++ and Go
programs. It employs runtime instrumentation and happensbefore analysis to identify potential data races. Inspector,
designed for Java programs, utilizes dynamic analysis techniques, including happens-before relations and synchronization tracking, to identify race conditions during program
execution. Microsoft Research’s CHESS \cite{chess} is a dynamic
analysis tool that systematically explores thread schedules
to detect potential race conditions, deadlocks, and other
concurrency issues in concurrent programs. FastTrack \cite{fasttrack},
tailored for C/C++ concurrent programs, combines happensbefore analysis and lockset-based techniques for dynamic race
detection. Lastly, ConTest \cite{contest} focuses on multi-threaded Java
programs, employing dynamic analysis and formal methods to
offer precise race detection and in-depth analysis of execution
schedules.
These tools provide valuable assistance in identifying and mitigating concurrency-related problems during runtime.In summary, these dynamic analysis tools offer diverse approaches to
detect and address concurrency bugs, ranging from data races
and atomicity violations to deadlocks and program crashes,
providing valuable assistance to programmers in improving
software reliability and robustness.

\subsection{Hybrid Analysis Tools}
In addition to the two primary types of concurrent bug detectors mentioned above, there exist additional strategies, as shown in Table\ref{table:hybrid}, that either blend static and dynamic methods together or combine one of the two ways with other approaches. Scholars have acknowledged the shortcomings of previous methods in identifying concurrent issues, especially those that depend on particular synchronization semantics, which may result in false reports of flaws.
To address these limitations, Lu et al. introduced AVIO \cite{lu2007avio}, a tool for detecting atomicity violations, which combines static and dynamic approaches. AVIO identifies AccessInterleaving invariants (AI invariants) that represent code
sections expected to execute atomically. Concurrency problems are identified if memory access interleavings break these invariants during runtime. 
Similarly, Shi et al. introduced definitionuse invariants (DefUse invariants)\cite{shi2010defuse} extracted from training runs
to dynamically detect violations and different types of program
bugs. Zhang et al. proposed a consequence-oriented approach,
focusing on a bug’s lifecycle stages, leading to improved
bug detection accuracy and coverage\cite{zhang2011conseq}. Kasikci et al.
presented RaceMob\cite{kasikci2013racemob},a data race detector that guarantees low runtime overhead and good accuracy by combining static and dynamic discovery.Deng et al. introduced Concurrent Function
Pair (CFP)\cite{deng2013efficient}, an interleaving-coverage metric, to improve
bug-detection efficiency in dynamic analysis. Additionally,
innovative tools like MUVI\cite{lu2007avio}, which detects inconsistent
updates and multi-variable concurrency bugs, and Portend+\cite{kasikci2015automated}, which have been designed to improve concurrency bug identification and analysis. It not only detects data races but also evaluates possible repercussions to classify them depending on severity. Helgrind\cite{helgrind}is a versatile race detection tool
within the Valgrind framework, employing a hybrid approach
that combines both dynamic and static analysis methods for
C/C++ programs. It effectively identifies race conditions by
utilizing lockset-based analysis to uncover potential data races.
Similarly, Eraser\cite{eraser} is another hybrid tool designed for multithreaded C/C++ programs, employing a combination of static
analysis and dynamic instrumentation to detect races. It relies
on lockset-based analysis and thread preemption techniques to
identify potential data race issues. RoadRunner\cite{roadrunner} focuses
on concurrent Java programs and employs a hybrid dynamic static analysis approach to uncover race conditions and other
concurrency-related problems. However, Saturn \cite{saturn} is designed specifically for multithreaded C/C++ programs. It uses a combination of dynamic symbolic execution and static analysis to identify any data races and provide accurate execution traces.
Finally, SWORD \cite{sword} operates in the realm of multithreaded Java
programs, employing a hybrid approach that integrates static
and dynamic analysis techniques. It conducts static analysis to
identify potential races and then employ dynamic analysis to
validate and refine the results, ensuring effective race condition
detection. These tools collectively leverage both static and
dynamic methods to overcome the limitations of each and
provide more effective bug detection and classification.
\begin{table*}[h]
\caption{Hybrid Detection Techniques}
\begin{center}
\label{table:hybrid}
\begin{tabular}{p{2cm}p{1cm}p{1.5cm}p{2cm}p{5cm}p{4cm}}

\textbf{Tool} & \textbf{Year} & \textbf{Author} & \textbf{PUT} & \textbf{Techniques used} & \textbf{Key Benefits} \\
\hline
PACER\cite{bond2010pacer} & 2010 & Bond et.al & Real world applications & Proportional Analysis & Efficiency, precision and versatility \\
\hline
RaceMob\cite{kasikci2013racemob} & 2013 & Kasikci et.al & Real world applications & Crowd sourcing + statistical significance & Always-on in production, Accuracy, Low overhead \\
\hline
ConSeq\cite{zhang2011conseq} & 2011 & Zhang et.al & C/C++ applications & Sequential replay & Increased bug detection coverage, lower false positive \\
\hline
MultiRace\cite{pozniansky2007multirace} & 2007 & Pozniansky et.al & C++ applications & Djit and Lockset & Lower overhead, Scalability \\
\hline
ColFinder\cite{wu2015collaborative} & 2015 & Wu. et. al & Program, written in C language & static program slicing +Thread scheduling & Reduce time of bug manifestation and overhead \\
\hline
SDRacer\cite{wang2020automatic} & 2020 & Wang et.al & Embedded programs written in C language & Static analysis+ Symbolic Execution & Automation, Accutacy and efficiency \\
\hline
HistLock+\cite{yang2018histlockc} & 2018 & Yang.et.al & C and C++ program & Lockset-based + History-based  & Completeness, Precision, Efficiency,Versatility \\
\hline
SVD\cite{xu2005serializability} & 2005 & Xu et.al & C programs & Lockset analysis + Happens-before analysis+ Atomic region analysis & Improved Reliability and lower false positive \\
\hline
Helgrind\cite{helgrind} & 2005 & Valgrind & C/C++ programs & Part of Valgrind tool suite+instrumenting program code & Ease of use, Accuracy \\
\hline
\end{tabular}
\end{center}
\end{table*}
\subsubsection{Fuzzers}
As discussed earlier one of the hybrid technique used for bug detection which is quite prominently used nowadays both for sequential and concurrent applications is fuzzing. Most fuzzing techniques for identifying concurrency bugs rely on static analysis to extract bug manifestation features from real-world concurrency issues. Fuzzing, as a method for deliberately exposing concurrency bugs, is considered more efficient compared to other approaches.
Race-Fuzzer\cite{racefuzz}, developed by Sen et. al in 2008, is designed to efficiently reproduce data races while keeping overhead to a minimum. The approach initially identifies potential races using a detection technique. It then employs a randomized thread scheduler to manipulate thread execution in a way that triggers actual data races from the previously identified potential ones. Race-Fuzzer is effective at distinguishing real data races, which have the potential to cause program exceptions, from other potential races. However, its effectiveness in exposing bugs depends heavily on the underlying data race detection tools, potentially missing some bugs due to limited coverage.
For detecting deadlocks, DeadlockFuzzer\cite{deadlockfuzz}, introduced by Joshi et al. in 2009, identifies real deadlocks in multithreaded programs. It begins by locating suspicious deadlocks using dynamic analysis.DeadlockFuzzer then uses a randomized scheduler to regulate thread scheduling in order to replicate these suspicious deadlocks, hence raising the probability of exposing them. It has been noted, meanwhile, that DeadlockFuzzer might have trouble accurately confirming the presence of a true deadlock.
ConLock\cite{Confuzz}, a mechanism that dynamically tests for deadlocks using constraint-based approaches, was invented by Cai et al. (2014) in order to address this problem. ConLock begins by examining a possible deadlock situation in order to derive a set of thread scheduling constraints. The rules guiding the relevant thread pairs' acquisition and release of locks during the possible deadlock are specified by these limitations. After then, the program is run within the limitations in an effort to produce a deadlock. ConLock labels a stalemate as false positive if it is not reproducible. 
Linux kernels using both lockless and lock-based synchronization methods for concurrency are more susceptible to bugs. Therefore, there have been multiple fuzzers developed for detecting concurrency bug in kernel programs.\cite{razzer}\cite{DDRACE}\cite{Krace}\cite{Conzzer}. 
Table\ref{table:fuzzer} gives a brief overview of the fuzzers developed for concurrency bug detection. 
\begin{table*}[h]
    \centering
    \caption{Concurrency Bug Detection Fuzzers}
    \label{table:fuzzer}
   \begin{tabular}{p{2cm}p{3cm}p{1.5cm}p{2cm}p{4cm}p{3cm}}
\hline
\textbf{Fuzzer} & \textbf{Author} & \textbf{Year }& \textbf{PUT }& \textbf{Contribution} & \textbf{Limitation} \\
\hline
ConFuzz\cite{Confuzz} & Padhiyar et.al & 2021 & OCaml Programs & Builds on AFL to generate inputs that maximize coverage of concurrent event-driven programs' non-deterministic state space. & Computationally expensive and can generate false positives \\
\hline
CalFuzzer\cite{calfuzz} & Koushik sen et.al & 2007 & C Programs & Uses the RAPOS algorithm to reduce sampling non-uniformity, leading to more efficient and effective testing. & needs to explicitly control the scheduler of the concurrent program to trigger concurrency bugs \\
\hline
Atomfuzzer\cite{atom} & Park et.al & 2008 & JAVA & Real-world flaws are effectively found by Java's randomized dynamic analysis technique, which alters the thread scheduler to discover atomicity violations with a high likelihood. & Computationally expensive, False Positive \\
\hline
Race Fuzzer\cite{racefuzz} & Koushik sen et.al & 2008 & JAVA & Propose a new approach that is based on the idea of directing the random testing process towards areas of the program that are more likely to contain race conditions. & False Positive, Scalibility, Overhead \\
\hline
Deadlockfuzzer\cite{deadlockfuzz} & Pallavi Joshi et.al & 2009 & JAVA & Uses the LDG to identify potential deadlocks and then use a random thread scheduler to create a potential deadlock & False Negative, False Positive, Overhead \\
\hline
Assetfuzzer\cite{Assetfuzz} & Zhifeng et.al & 2010 & C programs & The method prunes those interleavings that are free of violations and infers possible violations that do not show up in a particular execution.& False Negative, Overhead \\
\hline
Magicfuzzer\cite{Magic} & Yan Cai et.al & 2012 & C programs & Founded on a unique strategy that prunes lock dependencies repeatedly until none of them has an incoming or outgoing edge.& Overhead and False Positive \\
\hline
RAZZER\cite{razzer} & Dae R. Jeong et. al & 2019 & Linux Kernel & Combines the strengths of both stateless and stateful fuzzing to find kernel race bugs & False Positive, Scalibility, Overhead \\
\hline
DDRace\cite{DDRACE} & Yuan et.al & 2023 & Linux Kernel & Uses directed fuzzing to detect inputs that cause concurrency UAF vulnerabilities after it has identified target driver interfaces and race pairs. & False Negative, Overhead \\
\hline
KRACE\cite{Krace} & Meng Xu et.al & 2020 & Linux Kernel & Combines the strengths of both stateless and stateful fuzzing to find data race bugs in kernel file systems & False Negative, False Positive, Overhead \\
\hline
CONZZER\cite{Conzzer} & Zu-Ming Jiang et.al & 2022 & User and Kernel & Technique that uses knowledge of the program's context and direction of execution to generate more effective inputs for detecting data-races. & Complex, Computationally Expensive,cannot detect all data races \\
\hline
\end{tabular}

\end{table*}

\section{Discussion of detection methods and Mitigation techniques}
\begin{figure*}[h]
    \centering 
    \begin{subfigure}
         \centering
         \includegraphics[width=0.6\textwidth]{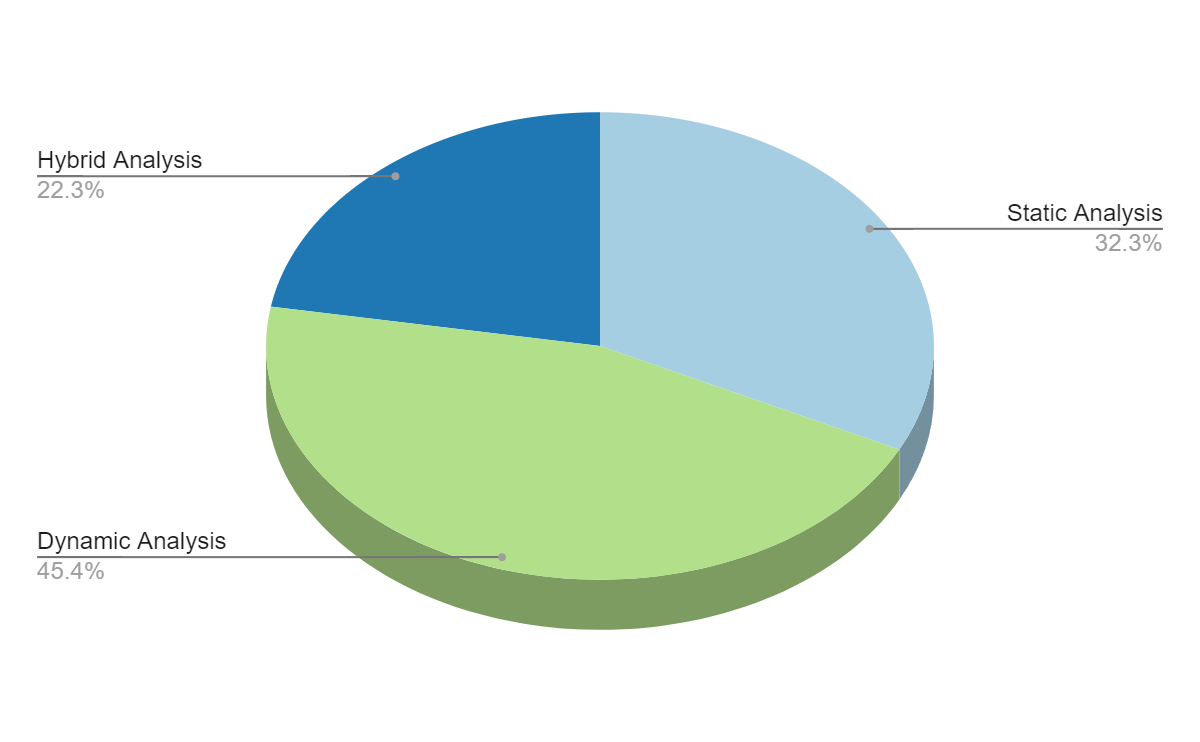}
     \end{subfigure}
     \begin{subfigure}
         \centering
         \includegraphics[width=0.6\textwidth]{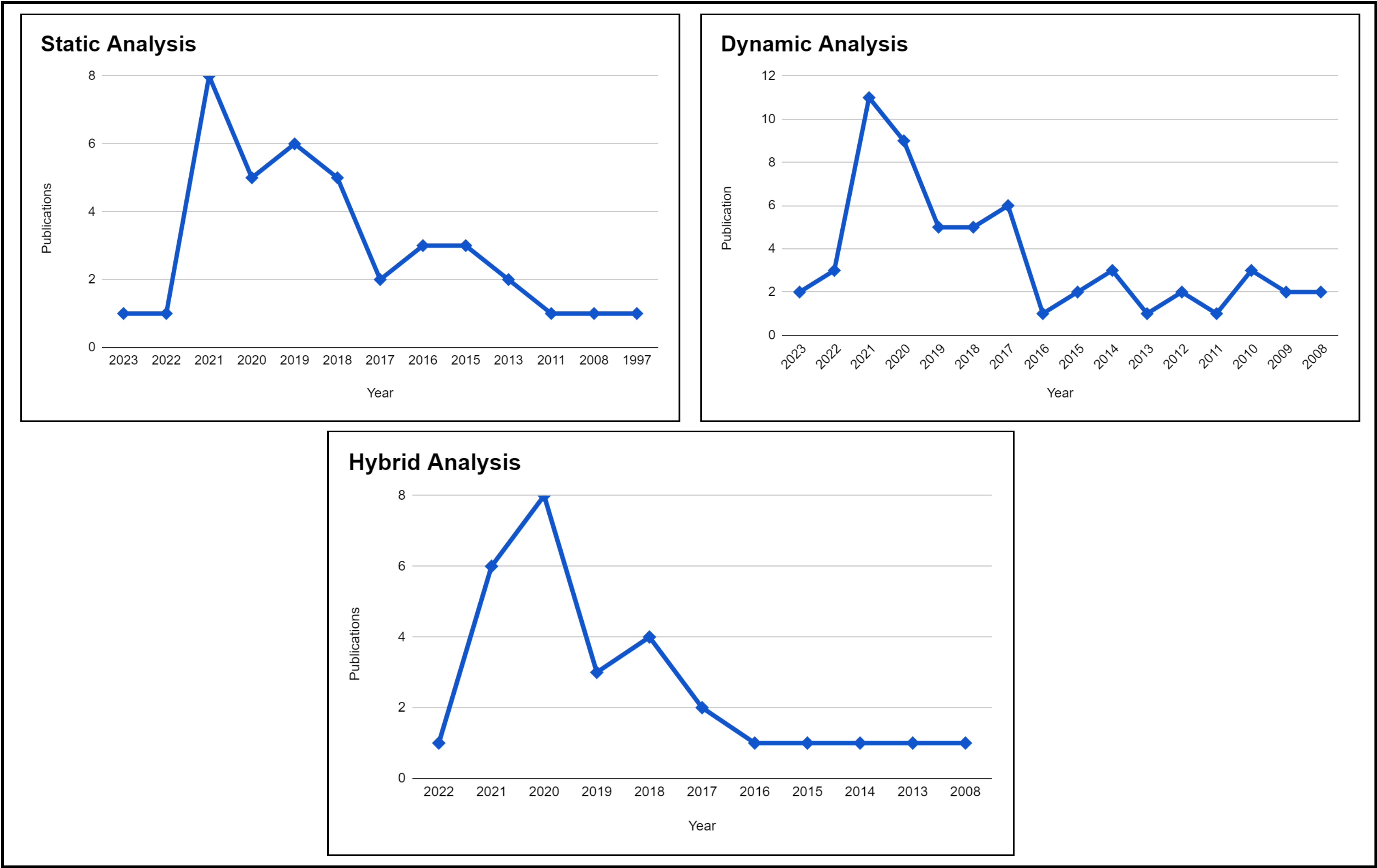}
     \end{subfigure}
    
    \caption{(a) Shows the distribution of techniques used in data race detection (b) Shows the statistical analysis of different detection techniques used over the years.}
    \label{fig:secondtwo}
\end{figure*}

The application analysis method employed in 32.3\% and 45.4\% of the analyzed studies, respectively, was static analysis and dynamic analysis. The method of hybrid analysis was employed by the remaining 22.3\%. A diagram showing this is shown in Fig\ref{fig:secondtwo}(a).As demonstrated in Figure\ref{fig:secondtwo}(b), the adoption of detecting techniques has also changed, impacting conventional procedures. Static procedures were one of the main methods for detection in the early years, but as the 20th century approaches, dynamic and hybrid techniques become more prominent. 
Because static analysis concentrates more on code features than code-level analysis, it may be more commonly used than code-level analysis. In addition, the cost of static analysis is less than that of the other two techniques. To run the source code, dynamic needs extra resources like emulators or actual hardware. However, as technology advanced, programs grew more complicated and were harder to solve using static analysis. Therefore, in order to lower the complexity and subsequently the overheads, formal and dynamic methods of symbolic execution were employed. 

Based on the review study, we also infer that the use of machine learning has been lower in the detection of data race bugs which can be a good direction to explore.
\subsection{Mitigation Techniques}
Once a vulnerability has been detected, several strategies can be used to mitigate it. These include locking and synchronization mechanisms, such as semaphores and mutexes, which can prevent race conditions and deadlocks. Other mitigation strategies include input validation, error checking, and memory protection mechanisms such as ASLR and DEP. Software developers can also use programming languages and libraries designed to reduce the risk of concurrency-related vulnerabilities, such as Rust and Go.
\section{Conclusion \& Future Directions}
We have conducted a comprehensive assessment of the most current and critical tools for detecting data race bugs. We have organized these tools into categories based on the specific types of techniques they have used for bug identification. This categorization can help researchers understand which types of techniques have been the focus of previous detectors and provide insights for developing new algorithms that can complement existing tools.
Furthermore, we have briefly discussed various techniques that have been used in recent times to enhance the effectiveness and efficiency of data race detection. These techniques include methods such as symbolic execution, fuzzing, guided coverage, proportional analysis, and static slicing. Researchers can leverage these techniques when designing new concurrency bug detectors.
A statistical study of several data race bug detectors is also included in our analysis, taking into account the different types of approaches that each one uses. Notably, the discovery of data race bugs frequently makes use of dynamic analytic approaches. We have proposed future study avenues based on our findings, which can help researchers working on data race detection.Additionally, we can conclude that the majority of currently available detectors experience false negatives if we take into account the accuracy, precision, performance, and application of the devices. These results might direct future research efforts aimed at reducing the drawbacks of data race detection.Overall, we expect to see significant advancement in the area of concurrency bug detection.
Our analysis has yielded valuable insights into these limitations and therefore our view of the present scenario and future scope can be seen as : 
\begin{itemize}
 \item Concurrency bugs are rarely discovered due to two primary factors: the immense and exponentially expanding interleaving possibilities as code size increases, and the fact that these bugs are typically hidden within uncommon and unique memory access interleaving patterns.
    \item Recent research has made progress in detecting and mitigating race conditions in concurrent execution but challenges remain. Balancing performance and security is crucial, and researchers are exploring dynamic analysis techniques while seeking more automated methods for identifying race conditions. Future research should focus on improved techniques, formal methods for correctness, and secure programming languages.
    \item Static detectors produce few false negatives but numerous false positives, posing challenges for debugging. Dynamic detectors can miss some concurrency bugs due to unexecuted code and may be inefficient. Research directions include reducing false negatives in dynamic detectors, addressing the accuracy-efficiency trade-off, and designing detectors to find real source code bugs and library-related issues
    \item Detectors often require manual annotations, limiting their use in complex programs. Researchers aim to automate these detectors.Some detectors generate excessive false positives, hindering bug identification and resolution. Improving accuracy enhances their applicability. A two-step detection process, akin to hybrid detectors, could be employed, with a focus on improving convenience.Most detectors are designed for C/C++ and Java, while languages like Objective-C and Python are also prevalent. There's a need for detectors compatible with a broader range of programming languages.
    \item Some dynamic detectors slow down program execution due to excessive instrumentation. Researchers can mitigate this by employing strategies like sampling or using virtual machines for monitoring.Dynamic detectors often execute programs redundantly, impacting efficiency. Strategies should be developed to eliminate redundant dynamic analyses.To make detectors suitable for production environments, researchers should work on hardware and algorithms to minimize runtime overhead.Deterministic execution techniques reduce non-determinism in large programs, improving software reliability and aiding data race bug detection. Predictive analysis can identify safe code, reducing the number of interleavings to be analyzed.
    \item Use of Formal methods such as model checking and Symbolic execution reduces the complexity of the programs and thus needs to be explored more in the area of guided symbolic execution.
    \item A promising research direction in the field of data race bugs, can be a detection solution that focuses on the collaboration between software and hardware. While software-only solutions are versatile and can be used on different platforms, they often slow down program performance. On the other hand, some systems try to make hardware changes to improve efficiency, but these modifications might not work well in a wide range of situations.
\end{itemize}

 \printbibliography

@article{dell-ht,
author = {Ali, Rizwan and Celebioglu, Onur and Hsieh, Jenwei and leng, tau},
journal= {dell high performace computing},
year = {2003},
month = {12},
pages = {},
title = {Intel Hyper-Threading Technology to Achieve Computational Efficiency HIGH-PERFORMANCE COMPUTING}
}

@INPROCEEDINGS{NASA,
  author={Saini, Subhash and Jin, Haoqiang and Hood, Robert and Barker, David and Mehrotra, Piyush and Biswas, Rupak},
  booktitle={2011 18th International Conference on High Performance Computing}, 
  title={The impact of hyper-threading on processor resource utilization in production applications}, 
  year={2011},
  volume={},
  number={},
  pages={1-10},
  doi={10.1109/HiPC.2011.6152743}}

@article{HT-L1,
  title={Performance Evaluation of Intel Broadwell Nodes Based Supercomputer Using Computational Fluid Dynamics and Climate Applications},
  author={Subhash Saini and Robert T. Hood},
  journal={2017 IEEE 19th International Conference on High Performance Computing and Communications Workshops (HPCCWS)},
  year={2017},
  pages={58-65}
}

@article{HTL2,
author = {Hassanein, Wessam and Rashid, Layali and Hammad, Moustafa},
year = {2008},
month = {04},
pages = {206-225},
title = {Analyzing the Effects of Hyperthreading on the Performance of Data Management Systems},
volume = {36},
journal = {International Journal of Parallel Programming},
doi = {10.1007/s10766-007-0066-x}
}

@article{htl3,
author = {Emmanuel, Okonta and Ajani, D},
journal= {West African Journal of Industrial and Academic Research},
year = {2019},
month = {07},
pages = {},
title = {Performance Evaluation of Hyper Threading Technology Architecture Using Microsoft Operating System Platform}
}

@article{racerx,
  title={RacerX: Effective, Static Detection of Race Conditions and Deadlocks},
  author={Engler, Dawson and Ashcraft, Kathryn},
  journal={ACM SIGOPS Operating Systems Review},
  volume={37},
  number={5},
  pages={237--252},
  year={2003},
  publisher={ACM}
}

@article{coverity,
  title={Using static analysis to find bugs},
  author={Chandra, Satish and Gopalakrishnan, Ganesh and Qadeer, Shuvendu and Rehof, Jakob},
  journal={IEEE Software},
  volume={25},
  number={5},
  pages={22--29},
  year={2008},
  publisher={IEEE}
}

@article{cognicrypt,
  title={CogniCrypt: supporting developers in using cryptography},
  author={Somorovsky, Juraj and Heiderich, Mario and Jensen, Meiko and Schwenk, J{\"o}rg},
  journal={IEEE Security \& Privacy},
  volume={14},
  number={4},
  pages={68--77},
  year={2016},
  publisher={IEEE}
}

@article{threadsafe,
  title={ThreadSafe: Static analysis for thread-locality},
  author={Flanagan, Cormac and Freund, Stephen N and Saxe, Joshua B},
  journal={ACM Transactions on Programming Languages and Systems (TOPLAS)},
  volume={32},
  number={4},
  pages={13},
  year={2010},
  publisher={ACM}
}

@article{polyspace,
  title={Static detection of concurrency defects},
  author={Berdine, Joshua and Calcagno, Cristiano and Cook, Byron and Distefano, Dino and O'Hearn, Peter W},
  journal={ACM SIGPLAN Notices},
  volume={43},
  number={1},
  pages={205--216},
  year={2008},
  publisher={ACM}
}

@inproceedings{threadsanitizer,
  title={ThreadSanitizer: data race detection in practice},
  author={Serebryany, Konstantin and Burrows, Alexander and Kozlov, Dmitry and Volovyk, Dmitriy and Potapenko, Anna},
  booktitle={Proceedings of the 2012 ACM SIGPLAN International Conference on Object-Oriented Programming, Systems, Languages, and Applications},
  pages={265--284},
  year={2012},
  organization={ACM}
}

@inproceedings{chess,
  title={CHESS: a systematic testing tool for concurrent programs},
  author={Musuvathi, Madanlal and Narayanasamy, Satish and Burckhardt, Sebastian},
  booktitle={Proceedings of the ACM SIGPLAN 2008 conference on Programming language design and implementation},
  pages={335--348},
  year={2008},
  organization={ACM}
}

@article{fasttrack,
  title={FastTrack: efficient and precise dynamic race detection},
  author={Flanagan, Cormac and Freund, Stephen N},
  journal={ACM SIGPLAN Notices},
  volume={39},
  number={6},
  pages={121--133},
  year={2004},
  publisher={ACM}
}

@inproceedings{contest,
  title={ConTest: Speculative testing for concurrent programs},
  author={Godefroid, Patrice and Klarlund, Nils and Sen, Koushik},
  booktitle={Proceedings of the 2005 ACM SIGPLAN conference on Programming language design and implementation},
  pages={52--61},
  year={2005},
  organization={ACM}
}

@inproceedings{helgrind,
  title={Valgrind: A framework for heavyweight dynamic binary instrumentation},
  author={Seward, Julian and Nethercote, Nicholas and Weidendorfer, Josef},
  booktitle={Proceedings of the 2005 ACM SIGPLAN conference on Programming language design and implementation},
  pages={89--100},
  year={2005},
  organization={ACM}
}

@inproceedings{eraser,
  title={Eraser: a dynamic data race detector for multithreaded programs},
  author={Savage, Stefan and Burrows, Mike and Nelson, Greg and Sobalvarro, Philip and Anderson, Thomas},
  booktitle={ACM SIGPLAN Notices},
  volume={29},
  number={6},
  pages={27--37},
  year={1994},
  organization={ACM}
}

@inproceedings{roadrunner,
  title={RoadRunner: Towards Automatic Checkpointing of Whole-System Computations},
  author={Cui, Zheng and Yang, Xuejun and Zhou, Yiying},
  booktitle={Proceedings of the 2013 USENIX Annual Technical Conference (USENIX ATC)},
  pages={179--190},
  year={2013},
  organization={USENIX Association}
}

@article{saturn,
  title={Saturn: efficient software error detection using relaxed dependences},
  author={Chen, Wenguang and Lin, Ying and Xue, Jingling},
  journal={IEEE Transactions on Software Engineering},
  volume={32},
  number={11},
  pages={873--890},
  year={2006},
  publisher={IEEE}
}

@inproceedings{sword,
  title={SWORD: a developer-oriented tool for detecting concurrency bugs},
  author={Yang, Chunhua and Su, Zhendong},
  booktitle={Proceedings of the 2008 ACM SIGPLAN conference on Programming language design and implementation},
  pages={45--56},
  year={2008},
  organization={ACM}
}

@book{hennessy2011computer,
  title={Computer Architecture: A Quantitative Approach},
  author={Hennessy, John L. and Patterson, David A.},
  year={2011},
  publisher={Elsevier}
}

@article{hwu2011simd,
  title={SIMD architecture and vectorization},
  author={Hwu, Wen-mei W.},
  journal={ACM SIGARCH Computer Architecture News},
  volume={39},
  number={1},
  pages={2--13},
  year={2011},
  publisher={ACM}
}

@article{li2017survey,
  title={A survey of FPGA-based accelerator systems},
  author={Li, Zhen and Zhang, Bo and Li, Xiaowei and Li, Yajuan},
  journal={ACM Computing Surveys (CSUR)},
  volume={50},
  number={3},
  pages={34},
  year={2017},
  publisher={ACM}
}

@book{jacob2011memory,
  title={Memory systems: cache, DRAM, disk},
  author={Jacob, Bruce and Ng, Spencer and Wang, David T.},
  year={2011},
  publisher={Morgan Kaufmann}
}

@article{flanagan2010threadsafe,
  title={ThreadSafe: Static analysis for thread-locality},
  author={Flanagan, Cormac and Freund, Stephen N and Saxe, Joshua B},
  journal={ACM Transactions on Programming Languages and Systems (TOPLAS)},
  volume={32},
  number={4},
  pages={13},
  year={2010},
  publisher={ACM}
}

@article{gupta,
  author = {Gupta, R. K.},
  title = {Heterogeneous Multiprocessor Architectures for Embedded Systems},
  journal = {Proceedings of the IEEE},
  volume = {89},
  number = {3},
  pages = {420--437},
  month = {3},
  year = {2001},
}

@article{kumar,
  author = {Kumar, S. and Raghavendra, D. and Kallianpur, L.},
  title = {A Survey of Asymmetric Multiprocessing in Modern SoCs},
  journal = {IEEE Design \& Test},
  volume = {33},
  number = {1},
  pages = {45--55},
  month = {2},
  year = {2016},
}

@inproceedings{lebel,
  author = {Lebel, D. and Levinsky, J. and Morin, D.},
  title = {Asymmetric Multiprocessing in Mobile and Embedded Systems},
  booktitle = {2013 IEEE 19th International Symposium on Asynchronous Circuits and Systems},
  pages = {49--56},
  month = {6},
  year = {2013},
}

@incollection{dutta,
  author = {Dutta, R. and Jha, N. K.},
  title = {Symmetric Multiprocessing},
  booktitle = {Handbook of Hardware/Software Codesign},
  publisher = {Springer},
  address = {Boston, MA},
  year = {2017},
}

@inbook{bosch,
  author = {Bosch, J.},
  title = {Symmetric Multiprocessing (SMP)},
  booktitle = {Encyclopedia of Parallel Computing},
  publisher = {Springer},
  address = {Boston, MA},
  year = {2011},
}

@book{dandamudi,
  author = {Dandamudi, S. P.},
  title = {Symmetric Multiprocessing},
  booktitle = {Guide to Assembly Language Programming in Linux},
  publisher = {Springer},
  address = {Boston, MA},
  year = {2005},
}

@book{silberschatz,
  author = {Silberschatz, A. and Galvin, P. B. and Gagne, G.},
  title = {Symmetric Multiprocessing},
  booktitle = {Operating System Concepts},
  publisher = {Wiley},
  year = {2018},
}

@article{industry5.0,
title = {Industry 5.0: Prospect and retrospect},
journal = {Journal of Manufacturing Systems},
volume = {65},
pages = {279-295},
year = {2022},
issn = {0278-6125},
doi = {https://doi.org/10.1016/j.jmsy.2022.09.017},
url = {https://www.sciencedirect.com/science/article/pii/S0278612522001662},
author = {Jiewu Leng and Weinan Sha and Baicun Wang and Pai Zheng and Cunbo Zhuang and Qiang Liu and Thorsten Wuest and Dimitris Mourtzis and Lihui Wang},
}

@article{VUl,
  title={Software vulnerability detection using deep neural networks: a survey},
  author={Lin, Guanjun and Wen, Sheng and Han, Qing-Long and Zhang, Jun and Xiang, Yang},
  journal={Proceedings of the IEEE},
  volume={108},
  number={10},
  pages={1825--1848},
  year={2020},
  publisher={IEEE}
}

@ARTICLE{vul2,
  author={Zeng, Peng and Lin, Guanjun and Pan, Lei and Tai, Yonghang and Zhang, Jun},
  journal={IEEE Access}, 
  title={Software Vulnerability Analysis and Discovery Using Deep Learning Techniques: A Survey}, 
  year={2020},
  volume={8},
  number={},
  pages={197158-197172},
  doi={10.1109/ACCESS.2020.3034766}}

@misc{CVEdetails,
    title        = {Vulnerabilities as Threats to IT Infrastructure},
    author       = {CVE details},
    year         = 2022,
    note         = {\url{https://www.cvedetails.com/browse-by-date.php} [Accessed: (June 28, 2023)]}
}

@article{industry5,
  title={Humans in Industry 5.0--A Paradigm Shift},
  author={Bauer, Wolfgang and Schlund, Stefan and Fr{\"a}mling, Kary and H{\"a}m{\"a}l{\"a}inen, Jukka},
  journal={Procedia CIRP},
  volume={110},
  pages={4--9},
  year={2022},
  publisher={Elsevier}
}

@article{Moore,
author = {Mollick, Ethan},
year = {2006},
month = {08},
pages = {62 - 75},
title = {Establishing Moore's Law},
volume = {28},
journal = {Annals of the History of Computing, IEEE},
doi = {10.1109/MAHC.2006.45}
}

@misc{Moore_intel,   
    title = {Cramming More Components onto Integrated Circuits
Moore’s Law},   
    url = {https://www.intel.com/content/www/us/en/history/virtual-vault/articles/moores-law.html},   
    author = {Intel},    
    note = {Accessed on July 10, 2023} 
}

@article{datarace,
author = {Hong, Shin and Kim, Moonzoo},
year = {2014},
month = {12},
pages = {},
title = {A survey of race bug detection techniques for multithreaded programmes},
volume = {25},
journal = {Software Testing, Verification and Reliability},
doi = {10.1002/stvr.1564}
}

@misc{therac,
  title = {{Archived webpage: Concurrent Programming (CPE9001) - Monash University}},
  author = {T. Gallagher},
  howpublished = {\url{https://web.archive.org/web/20071212183729/http://neptune.netcomp.monash.edu.au/cpe9001/assets/readings/www_uguelph_ca_~tgallagh_~tgallagh.html}},
  year = {2007},
  note = {Accessed on: July 31, 2023}
}

@inproceedings{chen1990dert,
  title = {{DERT}: A Software Tool for Detecting and Eliminating Racy Variables},
  author = {Chen, Peter M. and Hwang, Wen-Tsuen and Farber, David J.},
  booktitle = {Proceedings of the 1990 International Conference on Parallel Processing},
  volume = {1},
  pages = {375--378},
  year = {1990},
}

@inproceedings{castro2001empirical,
  title = {An Empirical Study of Operating Systems Errors},
  author = {Castro, Manuel},
  booktitle = {Proceedings of the 18th ACM Symposium on Operating Systems Principles (SOSP)},
  pages = {73--88},
  year = {2001},
}

@inproceedings{yang2011finding,
  title = {Finding and Understanding Bugs in C Compilers},
  author = {Yang, Sheng and Jiang, Yi and Cheng, Hao and Foster, Jeffrey S.},
  booktitle = {Proceedings of the 32nd ACM SIGPLAN Conference on Programming Language Design and Implementation (PLDI)},
  pages = {283--294},
  year = {2011},
}

@book{silberschatz2014operating,
  title = {Operating System Concepts},
  author = {Silberschatz, Abraham and Galvin, Peter B. and Gagne, Greg},
  edition = {9th},
  publisher = {Wiley},
  year = {2014},
}

@article{rodeh2013btrfs,
  title={BTRFS: The Linux B-tree filesystem},
  author={Rodeh, Ohad and Bacik, Josef and Mason, Chris},
  journal={ACM Transactions on Storage (TOS)},
  year={2013}
}

@misc{kernel-ext4-bugzilla,
    title        = {Kernel.org Bugzilla - ext4 bug entries},
    author       = {bugzilla.kernel},
    year         = 2022,
    note         = {\url{https://bugzilla.kernel.org/buglist.cgi?component=ext4} [Accessed: (22 august 2023)]}
}

@misc{kernel-btrfs-bugzilla,
    title        = {Kernel.org Bugzilla - Btrfs bug entries},
    author       = {bugzilla.kernel},
    year         = 2022,
    note         = {\url{https://bugzilla.kernel.org/buglist.cgi?component=btrfs} [Accessed: (22 august 2023)]}
}

@misc{kernel-bugzilla,
    title        = {Kernel.org Bugzilla},
    author       = {bugzilla.kernel},
    year         = 2022,
    note         = {\url{https://bugzilla.kernel.org} [Accessed: (22 august 2023)]}
}

@article{lu2014study,
  title={A study of Linux file system evolution},
  author={Lu, Lanyue and Arpaci-Dusseau, Andrea C. and Arpaci-Dusseau, Remzi H. and Lu, Shan},
  journal={Transactions on Storage},
  volume={10},
  number={1},
  pages={3:1--3:32},
  month={1},
  year={2014}
}

@inproceedings{huang2016evolutionary,
  title={An Evolutionary Study of Linux Memory Management for Fun and Profit},
  author={Huang, J. and Qureshi, M. K. and Schwan, K.},
  booktitle={Proceedings of the 2016 USENIX Annual Technical Conference (ATC)},
  address={Berkeley, CA, USA},
  month={6},
  year={2016},
  pages={465--478},
  isbn={978-1-931971-30-0}
}

@inproceedings{aghayev2019file,
  title={File Systems Unfit As Distributed Storage Backends: Lessons from 10 Years of Ceph Evolution},
  author={Aghayev, A. and Weil, S. and Kuchnik, M. and Nelson, M. and Ganger, G. R. and Amvrosiadis, G.},
  booktitle={Proceedings of the 27th ACM Symposium on Operating Systems Principles (SOSP)},
  address={Ontario, Canada},
  month={10},
  year={2019}
}

@inproceedings{min2016understanding,
  title={Understanding Manycore Scalability of File Systems},
  author={Min, C. and Kashyap, S. and Maass, S. and Kang, W. and Kim, T.},
  booktitle={Proceedings of the 2016 USENIX Annual Technical Conference (ATC)},
  address={Denver, CO},
  month={6},
  year={2016}
}

@misc{mitre-cve-2009-1235,
    title        = {CVE-2009-1235},
    author       = {MITRE Corporation},
    year         = 2009,
    note         = {\url{https://cve.mitre.org/cgibin/cvename.cgi?name=CVE-2009-1235} [Accessed: (22 august 2023)]}
}

@inproceedings{xu2019fuzzing,
  title={Fuzzing File Systems via Two-Dimensional Input Space Exploration},
  author={Xu, W. and Moon, H. and Kashyap, S. and Tseng, P.-N. and Kim, T.},
  booktitle={Proceedings of the 40th IEEE Symposium on Security and Privacy (Oakland)},
  address={San Francisco, CA},
  month={5},
  year={2019}
}

@misc{corbet2018unprivileged,
  title={Unprivileged filesystem mounts, 2018 edition},
  author={Corbet, J.},
  url={https://lwn.net/Articles/755593},
  year={2022}
}

@inproceedings{fonseca2014ski,
  title={SKI: Exposing Kernel Concurrency Bugs Through Systematic Schedule Exploration},
  author={Fonseca, P. and Rodrigues, R. and Brandenburg, B. B.},
  booktitle={Proceedings of the 11th USENIX Symposium on Operating Systems Design and Implementation (OSDI)},
  address={Broomfield, Colorado},
  month={10},
  year={2014}
}

@inproceedings{kim2019finding,
  title={Finding Semantic Bugs in File Systems with an Extensible Fuzzing Framework},
  author={Kim, S. and Xu, M. and Kashyap, S. and Yoon, J. and Xu, W. and Kim, T.},
  booktitle={Proceedings of the 27th ACM Symposium on Operating Systems Principles (SOSP)},
  address={Ontario, Canada},
  month={10},
  year={2019}
}

@misc{mitre-f2fs-cve,
    title        = {F2FS CVE entries},
    author       = {MITRE Corporation},
    year         = 2022,
    note         = {\url{http://cve.mitre.org/cgibin/cvekey.cgi?keyword=f2fs} [Accessed: (22 august 2023)]}
}

@InProceedings{Confuzz,
author="Padhiyar, Sumit
and Sivaramakrishnan, K. C.",
editor="Morales, Jos{\'e} F.
and Orchard, Dominic",
title="ConFuzz: Coverage-Guided Property Fuzzing for Event-Driven Programs",
booktitle="Practical Aspects of Declarative Languages",
year="2021",
publisher="Springer International Publishing",
address="Cham",
pages="127--144",
isbn="978-3-030-67438-0"
}

@inproceedings{calfuzz,
author = {Sen, Koushik},
title = {Effective Random Testing of Concurrent Programs},
year = {2007},
isbn = {9781595938824},
publisher = {Association for Computing Machinery},
address = {New York, NY, USA},
url = {https://doi.org/10.1145/1321631.1321679},
doi = {10.1145/1321631.1321679},
booktitle = {Proceedings of the 22nd IEEE/ACM International Conference on Automated Software Engineering},
pages = {323–332},
numpages = {10},
location = {Atlanta, Georgia, USA},
series = {ASE '07}
}

@inproceedings{atom,
author = {Park, Chang-Seo and Sen, Koushik},
title = {Randomized Active Atomicity Violation Detection in Concurrent Programs},
year = {2008},
isbn = {9781595939951},
publisher = {Association for Computing Machinery},
address = {New York, NY, USA},
url = {https://doi.org/10.1145/1453101.1453121},
doi = {10.1145/1453101.1453121},
booktitle = {Proceedings of the 16th ACM SIGSOFT International Symposium on Foundations of Software Engineering},
pages = {135–145},
numpages = {11},
location = {Atlanta, Georgia},
series = {SIGSOFT '08/FSE-16}
}

@inproceedings{racefuzz,
author = {Sen, Koushik},
title = {Race Directed Random Testing of Concurrent Programs},
year = {2008},
isbn = {9781595938602},
publisher = {Association for Computing Machinery},
address = {New York, NY, USA},
url = {https://doi.org/10.1145/1375581.1375584},
doi = {10.1145/1375581.1375584},
booktitle = {Proceedings of the 29th ACM SIGPLAN Conference on Programming Language Design and Implementation},
pages = {11–21},
numpages = {11},
location = {Tucson, AZ, USA},
series = {PLDI '08}
}

@inproceedings{deadlockfuzz,
author = {Joshi, Pallavi and Park, Chang-Seo and Sen, Koushik and Naik, Mayur},
title = {A Randomized Dynamic Program Analysis Technique for Detecting Real Deadlocks},
year = {2009},
isbn = {9781605583921},
publisher = {Association for Computing Machinery},
address = {New York, NY, USA},
url = {https://doi.org/10.1145/1542476.1542489},
doi = {10.1145/1542476.1542489},
booktitle = {Proceedings of the 30th ACM SIGPLAN Conference on Programming Language Design and Implementation},
pages = {110–120},
numpages = {11},
location = {Dublin, Ireland},
series = {PLDI '09}
}

@inproceedings{Assetfuzz,
author = {Lai, Zhifeng and Cheung, S. C. and Chan, W. K.},
title = {Detecting Atomic-Set Serializability Violations in Multithreaded Programs through Active Randomized Testing},
year = {2010},
isbn = {9781605587196},
publisher = {Association for Computing Machinery},
address = {New York, NY, USA},
url = {https://doi.org/10.1145/1806799.1806836},
doi = {10.1145/1806799.1806836},
booktitle = {Proceedings of the 32nd ACM/IEEE International Conference on Software Engineering - Volume 1},
pages = {235–244},
numpages = {10},
location = {Cape Town, South Africa},
series = {ICSE '10}
}

@inproceedings{Magic,
author = {Cai, Yan and Chan, W. K.},
title = {MagicFuzzer: Scalable Deadlock Detection for Large-Scale Applications},
year = {2012},
isbn = {9781467310673},
publisher = {IEEE Press},
booktitle = {Proceedings of the 34th International Conference on Software Engineering},
pages = {606–616},
numpages = {11},
location = {Zurich, Switzerland},
series = {ICSE '12}
}

@INPROCEEDINGS{razzer,
  author={Jeong, Dae R. and Kim, Kyungtae and Shivakumar, Basavesh and Lee, Byoungyoung and Shin, Insik},
  booktitle={2019 IEEE Symposium on Security and Privacy (SP)}, 
  title={Razzer: Finding Kernel Race Bugs through Fuzzing}, 
  year={2019},
  volume={},
  number={},
  pages={754-768},
  doi={10.1109/SP.2019.00017}}

@inproceedings{DDRACE,
  author       = {Ming Yuan and
                  Bodong Zhao and
                  Penghui Li and
                  Jiashuo Liang and
                  Xinhui Han and
                  Xiapu Luo and
                  Chao Zhang},
  editor       = {Joseph A. Calandrino and
                  Carmela Troncoso},
  title        = {DDRace: Finding Concurrency {UAF} Vulnerabilities in Linux Drivers
                  with Directed Fuzzing},
  booktitle    = {32nd {USENIX} Security Symposium, {USENIX} Security 2023, Anaheim,
                  CA, USA, August 9-11, 2023},
  publisher    = {{USENIX} Association},
  year         = {2023},
  url          = {https://www.usenix.org/conference/usenixsecurity23/presentation/yuan-ming},
  bibsource    = {dblp computer science bibliography, https://dblp.org}
}

@INPROCEEDINGS{Krace,
  author={Xu, Meng and Kashyap, Sanidhya and Zhao, Hanqing and Kim, Taesoo},
  booktitle={2020 IEEE Symposium on Security and Privacy (SP)}, 
  title={Krace: Data Race Fuzzing for Kernel File Systems}, 
  year={2020},
  volume={},
  number={},
  pages={1643-1660},
  doi={10.1109/SP40000.2020.00078}}

@inproceedings{Conzzer,
author       = {Zu{-}Ming Jiang and
Jia{-}Ju Bai and
Kangjie Lu and
Shi{-}Min Hu},
title        = {Context-Sensitive and Directional Concurrency Fuzzing for Data-Race
Detection},
booktitle    = {29th Annual Network and Distributed System Security Symposium, {NDSS}
2022, San Diego, California, USA, April 24-28, 2022},
publisher    = {The Internet Society},
year         = {2022},
url          = {https://www.ndss-symposium.org/ndss-paper/auto-draft-198/},
bibsource    = {dblp computer science bibliography, https://dblp.org}
}

@INPROCEEDINGS{lilibo,
  author={Bo, Lili and Meng, Xing and Sun, Xiaobing and Xia, Jingli and Wu, Xiaoxue},
  booktitle={2022 IEEE 22nd International Conference on Software Quality, Reliability and Security (QRS)}, 
  title={A Comprehensive Analysis of NVD Concurrency Vulnerabilities}, 
  year={2022},
  volume={},
  number={},
  pages={9-18},
  doi={10.1109/QRS57517.2022.00012}}

@article{abadi2006types,
  author = {Abadi, Martin and Flanagan, Cormac and Freund, Stephen N.},
  title = {Types for Safe Locking: Static Race Detection for Java},
  journal = {ACM Transactions on Programming Languages and Systems (TOPLAS)},
  volume = {28},
  pages = {207--255},
  year = {2006},
}

@inproceedings{relay,
  author = {Voung, Jason W. and Jhala, Ranjit and Lerner, Sorin},
  title = {Relay: Static Race Detection on Millions of Lines of Code},
  booktitle = {Joint Meeting of the European Software Engineering Conference and the ACM SIGSOFT International Symposium on Foundations of Software Engineering},
  pages = {205--214},
  year = {2007},
  address = {Dubrovnik, Croatia},
}

@article{vaziri,
  author = {Vaziri, Mandana and Tip, Frank and Dolby, Julian},
  title = {Associating Synchronization Constraints with Data in an Object-Oriented Language},
  journal = {ACM Sigplan Notices},
  volume = {41},
  pages = {334--345},
  year = {2006},
}

@article{smaragdakis2012sound,
  author = {Smaragdakis, Yannis and Evans, John and Sadowski, Caitlin and Yi, Jipeng and Flanagan, Cormac},
  title = {Sound Predictive Race Detection in Polynomial Time},
  journal = {ACM Sigplan Notices},
  volume = {47},
  pages = {387--400},
  year = {2012},
}

@article{choi2002efficient,
  author = {Choi, Jong-Deok and Lee, Kwangkeun and Loginov, Alexey and O’Callahan, Robert and Sarkar, Vivek and Sridharan, Manu},
  title = {Efficient and Precise Data Race Detection for Multithreaded Object-Oriented Programs},
  journal = {ACM Sigplan Notices},
  volume = {37},
  pages = {258--269},
  year = {2002},
}

@inproceedings{yu2005racetrack,
  author = {Yu, Yuan and Rodeheffer, Thomas and Chen, Wei},
  title = {Racetrack: Efficient Detection of Data Race Conditions via Adaptive Tracking},
  booktitle = {ACM SIGOPS Operating Systems Review},
  volume = {39},
  pages = {221--234},
  year = {2005},
  organization = {ACM},
}

@article{xu2005serializability,
  author = {Xu, Min and Bodík, Rastislav and Hill, Mark D.},
  title = {A Serializability Violation Detector for Shared-Memory Server Programs},
  journal = {ACM Sigplan Notices},
  volume = {40},
  pages = {1--14},
  year = {2005},
}

@inproceedings{ratanaworabhan2009detecting,
  author = {Ratanaworabhan, Paruj and Burtscher, Martin and Kirovski, Darko and Zorn, Benjamin and Nagpal, Rahul and Pattabiraman, Karthik},
  title = {Detecting and Tolerating Asymmetric Races},
  booktitle = {ACM SIGPLAN Notices},
  volume = {44},
  pages = {173--184},
  year = {2009},
  organization = {ACM},
}

@inproceedings{park2012unified,
  author = {Park, Soyeon and Vuduc, Richard and Harrold, Mary Jean},
  title = {A Unified Approach for Localizing Non-Deadlock Concurrency Bugs},
  booktitle = {2012 IEEE 5th International Conference on Software Testing, Verification and Validation},
  pages = {51--60},
  year = {2012},
  organization = {IEEE},
}

@inproceedings{park2010falcon,
  author = {Park, Soyeon and Vuduc, Richard W. and Harrold, Mary Jean},
  title = {Falcon: Fault Localization in Concurrent Programs},
  booktitle = {Proceedings of the 32nd ACM/IEEE International Conference on Software Engineering},
  volume = {1},
  pages = {245--254},
  year = {2010},
}

@article{jin2010instrumentation,
  author = {Jin, Guoliang and Thakur, Aditya and Liblit, Ben and Lu, Shan},
  title = {Instrumentation and Sampling Strategies for Cooperative Concurrency Bug Isolation},
  journal = {ACM Sigplan Notices},
  volume = {45},
  pages = {241--255},
  year = {2010},
}

@article{lucia2011isolating,
  author = {Lucia, Brandon and Wood, Brian P. and Ceze, Luis},
  title = {Isolating and Understanding Concurrency Errors Using Reconstructed Execution Fragments},
  journal = {ACM Sigplan Notices},
  volume = {46},
  pages = {378--388},
  year = {2011},
}

@article{zhang2010conmem,
  author = {Zhang, Wei and Sun, Chang and Lu, Shan},
  title = {ConMem: Detecting Severe Concurrency Bugs Through an Effect-Oriented Approach},
  journal = {ACM SIGARCH Computer Architecture News},
  volume = {38},
  pages = {179--192},
  year = {2010},
}

@article{zhang2016lightweight,
  author = {Zhang, Min and Wu, Yongming and Shan, Lu and Qi, Sheng and Ren, Jingyu and Zheng, Weimin},
  title = {A Lightweight System for Detecting and Tolerating Concurrency Bugs},
  journal = {IEEE Transactions on Software Engineering},
  volume = {42},
  number = {10},
  pages = {899--917},
  year = {2016},
}

@inproceedings{cai2012magicfuzzer,
  author = {Cai, Yifei and Chan, Wing-Kwong},
  title = {MagicFuzzer: Scalable Deadlock Detection for Large-Scale Applications},
  booktitle = {International Conference on Software Engineering},
  pages = {606--616},
  year = {2012},
}

@article{cai2014magiclock,
  author = {Cai, Yifei and Chan, Wing-Kwong},
  title = {MagicLock: Scalable Detection of Potential Deadlocks in Large-Scale Multithreaded Programs},
  journal = {IEEE Transactions on Software Engineering},
  volume = {40},
  pages = {266--281},
  year = {2014},
}

@article{lu2007avio,
  title={Avio: Detecting Atomicity Violations via Access-Interleaving Invariants},
  author={Lu, Shan and Tucek, Joseph and Qin, Feng and Zhou, Yuanyuan},
  journal={IEEE Micro},
  volume={27},
  pages={26--35},
  year={2007}
}

@article{shi2010defuse,
  title={Do I Use the Wrong Definition?: Defuse: Definition-Use Invariants for Detecting Concurrency and Sequential Bugs},
  author={Shi, Ying and Park, Soyeon and Yin, Zhihao and Lu, Shan and Zhou, Yuanyuan and Chen, Wei and Zheng, Wei},
  journal={ACM Sigplan Notices},
  volume={45},
  pages={160--174},
  year={2010}
}

@article{zhang2011conseq,
  title={Conseq: Detecting Concurrency Bugs Through Sequential Errors},
  author={Zhang, Wenguang and Lim, Jipeng and Olichandran, Rajeev and Scherpelz, Johann and Jin, Guoliang and Lu, Shan and Reps, Thomas},
  journal={ACM Sigplan Notices},
  volume={39},
  pages={251--264},
  year={2011}
}

@inproceedings{kasikci2013racemob,
  title={Racemob: Crowdsourced Data Race Detection},
  author={Kasikci, Baris and Zamfir, Cristian and Candea, George},
  booktitle={Twenty-Fourth ACM Symposium on Operating Systems Principles},
  pages={406--422},
  year={2013}
}

@article{deng2013efficient,
  title={Efficient Concurrency-Bug Detection Across Inputs},
  author={Deng, Dongdong and Zhang, Wenguang and Lu, Shan},
  journal={ACM Sigplan Notices},
  volume={48},
  pages={785--802},
  year={2013}
}

@article{kasikci2015automated,
  title={Automated Classification of Data Races Under Both Strong and Weak Memory Models},
  author={Kasikci, Baris and Zamfir, Cristian and Candea, George},
  journal={ACM Transactions on Programming Languages \& Systems},
  volume={37},
  pages={1--44},
  year={2015}
}

@book{tanenbaum2014modern,
  title={Modern Operating Systems},
  author={Tanenbaum, Andrew S. and Bos, Herbert},
  year={2014},
  edition={4th},
  publisher={Pearson}
}

@book{butenhof1997programming,
  title={Programming with POSIX Threads},
  author={Butenhof, David R.},
  year={1997},
  publisher={Addison-Wesley}
}

@book{silberschatz2018operating,
  title={Operating System Concepts},
  author={Silberschatz, Abraham and Galvin, Peter B. and Gagne, Greg},
  year={2018},
  edition={10th},
  publisher={Wiley}
}

@book{lewis1998threads,
  title={Threads Primer: A Guide to Multithreaded Programming},
  author={Lewis, J. P. and Berg, Mitchell},
  year={1998},
  publisher={Prentice Hall}
}

@article{harris2001pragmatic,
  title={A pragmatic implementation of non-blocking linked-lists},
  author={Harris, Timothy L.},
  journal={Distributed Computing},
  volume={14},
  number={3},
  pages={195--212},
  year={2001}
}

@article{moir2007using,
  title={Using elimination to implement scalable and lock-free FIFO queues},
  author={Moir, Mark and Nussbaum, Daniel and Shalev, Ori},
  journal={ACM Transactions on Programming Languages and Systems (TOPLAS)},
  volume={29},
  number={3},
  pages={1--34},
  year={2007}
}

@book{herlihy2008art,
  title={The Art of Multiprocessor Programming},
  author={Herlihy, Maurice and Shavit, Nir},
  year={2008},
  publisher={Morgan Kaufmann}
}

@misc{top500,
  author = {{TOP500.Org}},
  title = {Top 500 Supercomputer Sites},
  howpublished = {\url{http://www.top500.org/}},
  year = {2019},
  note = {Retrieved May 8, 2019},
}

@techreport{ieee2018,
  author = {Working Group for POSIX},
  title = {IEEE Standard for Information Technology--Portable Operating System Interface (POSIX(R)) Base Specifications, Issue 7. IEEE Std 1003.1-2017 (Rev. of IEEE Std 1003.1-2008)},
  year = {2018},
  month = {1},
  institution = {IEEE},
  pages = {1--3951},
  doi = {10.1109/IEEESTD.2018.8277153},
}

@article{deeprace,
  author       = {Ali Tehrani Jamsaz and
                  Mohammed Khaleel and
                  Reza Akbari and
                  Ali Jannesari},
  title        = {DeepRace: Finding Data Race Bugs via Deep Learning},
  journal      = {CoRR},
  volume       = {abs/1907.07110},
  year         = {2019},
  url          = {http://arxiv.org/abs/1907.07110},
  eprinttype    = {arXiv},
  eprint       = {1907.07110},
  bibsource    = {dblp computer science bibliography, https://dblp.org}
}

@inproceedings{erigone,
  author    = "M. Ben-Ari",
  title     = "Tool Presentation: Teaching Concurrency and Model Checking",
  booktitle = "Proceedings of the International SPIN Workshop on Model Checking Software",
  year      = 2009,
  pages     = "6-11",
  publisher = "Springer",
  address   = "Cham, Switzerland",
  doi       = "10.1007/978-3-642-02652-2\_5"
}

@article{Raceview,
  title={Advancing data race investigation and classification through visualization},
  author={Nikolaos Koutsopoulos and Mandy Northover and Timm Felden and Martin Wittiger},
  journal={2015 IEEE 3rd Working Conference on Software Visualization (VISSOFT)},
  year={2015},
  pages={200-204},
  url={https://api.semanticscholar.org/CorpusID:14574499}
}

@inproceedings{DRFRAME,
  author    = "D. Qi and N. Gu and J. Su",
  title     = "Detecting Data Race in Network Applications Using Static Analysis",
  booktitle = "Proceedings of the International Conference on Networking and Network Applications (NaNA)",
  month     = "10",
  year      = 2019,
  pages     = "313-318",
  doi       = "10.1109/NaNA.2019.00061"
}

@inproceedings{concurrentcfg,
  author    = "V. Kahlon and N. Sinha and E. Kruus and Y. Zhang",
  title     = "Static Data Race Detection for Concurrent Programs with Asynchronous Calls",
  booktitle = "Proceedings of the 7th Joint Meeting of the European Software Engineering Conference and the ACM SIGSOFT Symposium on the Foundations of Software Engineering (ESEC/FSE)",
  year      = 2009,
  pages     = "13-22",
  doi       = "10.1145/1595696.1595701"
}

@inproceedings{chord,
  author    = "M. Naik and A. Aiken and J. Whaley",
  title     = "Effective Static Race Detection for Java",
  booktitle = "Proceedings of the ACM SIGPLAN Conference on Programming Language Design and Implementation (PLDI)",
  year      = 2006,
  pages     = "308-319",
  publisher = "Association for Computing Machinery",
  address   = "New York, NY, USA",
  doi       = "10.1145/1133981.1134018"
}

@article{locksmith,
  author    = "P. Pratikakis and J. S. Foster and M. Hicks",
  title     = "LOCKSMITH: Practical Static Race Detection for C",
  journal   = "ACM Transactions on Programming Languages and Systems (TOPLAS)",
  volume    = 33,
  number    = 1,
  pages     = "1-55",
  month     = "1",
  year      = 2011,
  doi       = "10.1145/1889997.1890000"
}

@inproceedings{kiss,
  author    = "S. Qadeer and D. Wu",
  title     = "KISS: Keep It Simple and Sequential",
  booktitle = "Proceedings of the 25th Annual ACM SIGPLAN Conference on Programming Language Design and Implementation",
  year      = 2004,
  pages     = "14-24",
  publisher = "ACM",
  address   = "New York"
}

@inproceedings{EPAJ,
  author    = "R. Agarwal and A. Sasturkar and L. Wang and et al.",
  title     = "Optimized Run-time Race Detection and Atomicity Checking Using Partial Discovered Types",
  booktitle = "Proceedings of the 20th IEEE/ACM International Conference on Automated Software Engineering",
  year      = 2005,
  pages     = "233-242",
  publisher = "ACM",
  address   = "New York"
}

@inproceedings{cobe,
  author    = "V. Kahlon and N. Sinha and E. Kruus and et al.",
  title     = "Static Data Race Detection for Concurrent Programs with Asynchronous Calls",
  booktitle = "Proceedings of the 7th Joint Meeting of the European Software Engineering Conference and the ACM SIGSOFT Symposium on the Foundations of Software Engineering",
  year      = 2009,
  pages     = "13-22",
  publisher = "ACM",
  address   = "New York"
}

@inproceedings{iteracer,
  author    = "C. Radoi and D. Dig",
  title     = "Practical Static Race Detection for Java Parallel Loops",
  booktitle = "Proceedings of the International Symposium on Software Testing and Analysis",
  year      = 2013,
  pages     = "178-190",
  publisher = "ACM",
  address   = "New York"
}

@inproceedings{coderrect,
  author    = "Bozhen Liu* and Peiming Liu* and Yanze Li and Chia-Che Tsai and Dilma Da Silva and Jeff Huang",
  title     = "When Threads Meet Events: Efficient and Precise Static Race Detection with Origins",
  booktitle = "Proceedings of the 42nd ACM SIGPLAN International Conference on Programming Language Design and Implementation (PLDI '21)",
  month     = "6",
  year      = 2021,
  address   = "Virtual, Canada",
  publisher = "ACM",
  pages     = "15 pages",
  doi       = "10.1145/3453483.3454073",
  url       = "https://doi.org/10.1145/3453483.3454073"
}

@article{openrace,
  title     = "OpenRace: An Open Source Framework for Statically Detecting Data Races",
  author    = "Bradley Swain and Bozhen Liu and Peiming Liu and Yanze Li and Addison Crump and Rohan Khera and Jeff Huang",
  journal   = "Coderrect Inc.",
  address   = "College Station, TX",
  year      = {2021},  
  note      = {Emails: brad@coderrect.com, bozhen.liu@tamu.edu, peiming@coderrect.com, yanze@coderrect.com, addison.crump@coderrect.com, rohan.khera@coderrect.com, jeff@coderrect.com",}
}

@article{LLOV,
author = {Bora, Utpal and Das, Santanu and Kukreja, Pankaj and Joshi, Saurabh and Upadrasta, Ramakrishna and Rajopadhye, Sanjay},
title = {LLOV: A Fast Static Data-Race Checker for OpenMP Programs},
year = {2020},
issue_date = {December 2020},
publisher = {Association for Computing Machinery},
address = {New York, NY, USA},
volume = {17},
number = {4},
issn = {1544-3566},
url = {https://doi.org/10.1145/3418597},
doi = {10.1145/3418597},
journal = {ACM Trans. Archit. Code Optim.},
month = dec,
articleno = {35},
numpages = {26}
}

@inproceedings{ompracer,
  title     = "OMPRacer: A Scalable and Precise Static Race Detector for OpenMP Programs",
  author    = "Bradley Swain and Yanze Li and Peiming Liu and Ignacio Laguna and Giorgis Georgakoudis and Jeff Huang",
  booktitle = "SC20, November 9-19, 2020, Is Everywhere We Are",
  organization = "IEEE",
  year      = 2020,
  isbn      = "978-1-7281-9998-6",
  price     = "$31.00",
  address   = "College Station, TX and Livermore, CA",
  note      = "Affiliations and emails: Computer Science and Engineering, Texas A\&M University, College Station, TX, {brad, yanzeli, peiming}@tamu.edu, jeff@cse.tamu.edu; Center for Applied Scientific Computing, Lawrence Livermore National Laboratory, Livermore, CA, {ilaguna, georgakoudis1}@llnl.gov;0 Coderrect Inc, College Station, TX, brad@coderrect.com"
}

@inproceedings{GOBLINT,
author = {Vojdani, Vesal and Apinis, Kalmer and R\~{o}tov, Vootele and Seidl, Helmut and Vene, Varmo and Vogler, Ralf},
title = {Static Race Detection for Device Drivers: The Goblint Approach},
year = {2016},
isbn = {9781450338455},
publisher = {Association for Computing Machinery},
address = {New York, NY, USA},
url = {https://doi.org/10.1145/2970276.2970337},
doi = {10.1145/2970276.2970337},
booktitle = {Proceedings of the 31st IEEE/ACM International Conference on Automated Software Engineering},
pages = {391–402},
numpages = {12},
location = {Singapore, Singapore},
series = {ASE '16}
}

@inproceedings{bond2010pacer,
  author = {M. D. Bond and K. E. Coons and K. S. McKinley},
  title = {PACER: Proportional Detection of Data Races},
  booktitle = {Proceedings of the 31st Annual ACM SIGPLAN Conference on Programming Language Design and Implementation},
  year = {2010},
  pages = {255--268},
  organization = {ACM},
  address = {New York}
}

@article{pozniansky2007multirace,
  author = {E. Pozniansky and A. Schuster},
  title = {MultiRace: Efficient On-the-Fly Data Race Detection in Multithreaded C++ Programs},
  journal = {Concurrent Computation: Practice and Experience},
  year = {2007},
  volume = {19},
  pages = {327--340}
}

@article{wu2015collaborative,
  author = {Z. D. Wu and K. Lu and X. P. Wang and et al.},
  title = {Collaborative Technique for Concurrency Bug Detection},
  journal = {International Journal of Parallel Programming},
  year = {2015},
  volume = {43},
  pages = {260--285}
}

@article{wang2020automatic,
  author = {Y. Wang and F. Gao and L. Wang and T. Yu and J. Zhao and X. Li},
  title = {Automatic Detection, Validation, and Repair of Race Conditions in Interrupt-Driven Embedded Software},
  journal = {IEEE Transactions on Software Engineering},
  year = {2020},
  month = {4},
  note = {Early Access},
  doi = {10.1109/TSE.2020.2989171}
}

@article{yang2018histlockc,
  author = {J. Yang and B. Jiang and W. K. Chan},
  title = {HistLockC: Precise memory access maintenance without lockset comparison for complete hybrid data race detection},
  journal = {IEEE Transactions on Reliability},
  volume = {67},
  number = {3},
  pages = {786--801},
  month = {9},
  year = {2018},
  doi = {10.1109/TR.2018.2832226}
}

@article{key,
author = {Voronenko, Yevgen},
year = {2012},
month = {05},
pages = {},
title = {Library Generation For Linear Transforms Dissertation}
}


\end{document}